\documentclass[paper]{JHEP3}
\pdfoutput=1
\usepackage{epsfig}
\usepackage{amsmath}
\usepackage{amssymb}
\usepackage{amsbsy}
\usepackage{bbm}
\usepackage{enumerate}
\usepackage{url}
\usepackage{xspace}

\newlength{\wfig}
\newlength{\hfig}
\newlength{\hfigs}
\newcommand{\tmop}[1]{\ensuremath{{\rm #1}}}

\renewcommand{\H}{{\tt H}}
\newcommand\HJ{{\tt HJ}}
\newcommand\W{{\tt W}}
\newcommand\Z{{\tt Z}}
\newcommand\ZJ{{\tt ZJ}}
\newcommand\WJ{{\tt WJ}}
\newcommand\HJJ{{\tt HJJ}}

\newcommand\HWZ{{\tt H/W/Z}}
\newcommand\WZ{{\tt W/Z }}
\newcommand\B{{\tt B}}
\newcommand\BJ{{\tt BJ}}
\newcommand\BJJ{{\tt BJJ}}
\newcommand\yB{y_{\rm\scriptscriptstyle B}}

\newcommand\MH{M_{\scriptscriptstyle H}}

\def\beq{\begin{equation}}
\def\beqn{\begin{eqnarray}}
\def\eeq{\end{equation}}
\def\eeqn{\end{eqnarray}}

\def\half{\frac{1}{2}}

\newcommand\NNLOPS{NNLO+PS}
\newcommand\NLOPS{NLO+PS}

\newcommand\PYTHIA{{\tt PYTHIA}\xspace}

\newcommand\KRA{K_{\scriptscriptstyle \rm R}}
\newcommand\KFA{K_{\scriptscriptstyle \rm F}}

\newcommand\LambdaQCD{\Lambda_{\scriptscriptstyle \rm QCD}}

\newcommand\muf{\mu_{\sss\rm F}}
\newcommand\mur{\mu_{\sss\rm R}}

\def\lq{\left[} 
\def\rq{\right]} 
\def\rg{\right\}} 
\def\lg{\left\{} 
\def\({\left(} 
\def\){\right)}

\newcommand\sss{\mathchoice%
{\displaystyle}%
{\scriptstyle}%
{\scriptscriptstyle}%
{\scriptscriptstyle}%
}

\newdimen\hbigcirc
\newdimen\wbigcirc

\newdimen\figwidth

\figwidth=\textwidth

\newcommand\as{\alpha_{\sss\rm S}}

\newcommand\pt{p_{\sss \rm T}}

\newcommand\pTKH{q_{\sss \rm T}}

\newcommand\qt{q_{\sss\rm T}}
\newcommand\qT{q_{\sss\rm T}}

\newcommand\kT{k_{\sss\rm T}}

\newcommand\MCatNLO{{\tt MC@NLO}}

\newcommand\CA{C_{\sss\rm A}}

\newcommand\CF{C_{\sss\rm F}}

\newcommand\POWHEG{{\tt POWHEG}}
\newcommand\POWHEGBOX{{\tt POWHEG BOX}}

\newcommand\MINLO{{\tt MiNLO}}
\newcommand\MiNLO{{\tt MiNLO}}

\newcommand\ttilde{\raise.17ex\hbox{$\scriptstyle\mathtt{\sim}$}}

\def\ord#1{{\cal O}\(#1\)}

\newcommand{\mathd}{\mathrm{d}}

\newcommand{\Beta}{\mathrm{B}}

\newcount\minutes 
\newcount\scratch 
\def\timestamp{%
\scratch=\time 
\divide\scratch by 60 
\edef\hours{\the\scratch} 
\multiply\scratch by 60 
\minutes=\time 
\advance\minutes by -\scratch 
---$\,$\hours:\null 
\ifnum\minutes< 10 0\fi 
\the\minutes}

\preprint{CERN-PH-TH/2012-356\\LPN12-140, MCnet-12-15\\OUTP-13-04P}

\title{{Merging H/W/Z + 0 and 1 jet at NLO with no merging scale:
    a path to parton shower + NNLO matching}}

\author{Keith Hamilton\thanks{On leave from University College London.}\\
  Theory Division, CERN, CH--1211, Geneva 23, Switzerland\\
  E-mail: \email{keith.hamilton@cern.ch}
}
\author{Paolo Nason\\
  INFN, Sezione di Milano Bicocca, Piazza della Scienza 3, 20126 Milan, Italy\\
  E-mail: \email{paolo.nason@mib.infn.it}
}

\author{Carlo Oleari\\
  Universit\`a di Milano-Bicocca and INFN, Sezione di Milano-Bicocca\\
  Piazza della Scienza 3, 20126 Milan, Italy\\
  E-mail: \email{Carlo.Oleari@mib.infn.it}}

\author{Giulia Zanderighi\\
  Rudolf Peierls Centre for Theoretical Physics, 1 Keble Road, University of Oxford, UK\\
  E-mail: \email{g.zanderighi1@physics.ox.ac.uk}
}

\abstract{We consider the \POWHEG{} generator for a \HWZ{} boson plus one
  jet, augmented with the recently proposed \MINLO{} method for the choice of
  scales and the inclusion of Sudakov form factors.  Within this framework,
  the generator covers all the transverse-momentum region of the \HWZ{}
  boson, i.e.~no generation cuts are needed to obtain a finite result. By
  construction, the generator achieves NLO accuracy for distributions
  involving a finite (and relatively large) transverse momentum of the
  boson. We examine the conditions under which also the totally inclusive
  distributions (e.g.~the boson rapidity distribution) achieve NLO
  accuracy. We find that a minimal modification of the \MINLO{} prescription
  is sufficient to achieve such accuracy. We thus construct a NLO generator
  for \HWZ{} boson plus one jet production such that it smoothly merges into
  a NLO single boson production in the small transverse-momentum region. We
  notice that, by simply reweighting the boson rapidity distribution to NNLO
  predictions, we achieve a NNLO accurate generator matched to a shower. The
  approach applies to all production processes involving a colorless massive
  system plus one jet. We discuss how it may be extended to general
  processes.}

\begin{document}

\section{Introduction}
In recent times, next-to-leading order parton shower (\NLOPS{})
matching techniques have been developed  and realised as practical
simulation tools ~\cite{Frixione:2002ik,Nason:2004rx,Frixione:2007vw}.
The programs and methods underlying them have matured to the point
where they are routinely used in LHC data analysis.

\NLOPS{} methods can also be applied to processes involving associated jet
production at leading order.  In the case of Higgs boson production, for
example, there exist \POWHEG{} simulations for inclusive production of a
Higgs boson (the~\H{} generator from now on)~\cite{Alioli:2008tz}, Higgs
boson production with an associated jet~(\HJ{}), and also for the case of two
associated jets~(\HJJ{})~\cite{Campbell:2012am}.  These simulations overlap
in their population of phase space. However, the relative accuracies of each
one in the various regions is complementary. Specifically, the \H{} generator
yields NLO accurate inclusive Higgs boson distributions, describing the
radiation of a single jet with leading-order~(LO) accuracy and the radiation
of more than one jet with the accuracy of the shower to which it is
interfaced (the collinear, leading-log, approximation). On the other hand,
the \HJ{} generator gives NLO precision for observables inclusive in the
production of a Higgs boson plus one jet; it has only LO quality if we
require two jets, with further radiation generated in the parton shower
approximation. The \HJ{} generator cannot be used to describe inclusive jet
cross sections, since the NLO calculation on which it is based requires the
presence of at least one jet. Similar considerations hold for the \HJJ{}
generator.

It is natural to ask whether one can merge the \H{}, \HJ{} and
\HJJ{} simulations, in such a way that all classes of observables
have NLO accuracy in the end results. That is to say, one would
ideally like to have a unified \NLOPS{} simulation, or
simulation output, yielding inclusive Higgs boson distributions
accurate at the NLO level, $\as^3$,
and, at the same time, distributions involving the Higgs boson plus one jet
accurate at order $\as^4$.

Several collaborations have addressed this merging
problem~\cite{Lavesson:2008ah, Alioli:2011nr, Hoeche:2012yf, Gehrmann:2012yg,
  Frederix:2012ps, Alioli:2012fc, Platzer:2012bs, Lonnblad:2012ix}. They all
typically separate the output of each component simulation (the equivalent of
\H{}, \HJ{} or \HJJ{}) according to the jet multiplicity of the events it
produces, discarding those having a multiplicity for which the generator does
not possess the relevant NLO corrections. Having processed the output of each
simulation in this way, the event samples are joined to give an inclusive
sample. In a nutshell, each generator can therefore be regarded as
contributing a single exclusive jet bin to the final inclusive sample, the
magnitude of each bin being predominantly determined by the jet resolution
scale used in performing the merging, the so-called merging scale. The
merging scale is an unphysical parameter, and the dependence on it is,
rightly, well studied in the implementations of the merging algorithms cited
above, with different approaches invoking different means to mitigate the
dependence on it. Relatedly, we point out that in all of the practical
applications of NLO merging to date, the jet algorithms employed do not
correspond to those currently used in LHC experimental analyses, albeit for
good theoretical reasons: in particular, to lessen the dependence on the
merging scale through the resummation of spurious large logarithmic
corrections associated with it.

In all cases the choice of the merging scale poses a dilemma: if the merging
scale is too low, the sample is dominated by the higher-multiplicity
generators, while if it is chosen too large, one loses their benefit, since
one is forced to describe relatively hard jets only with tree-level accuracy,
or, worse still, with the parton-shower approximation.

To further clarify the situation consider again Higgs boson production.  As
stated above, the \H{} generator yields ${\cal O}(\as^3)$ accuracy for
inclusive quantities, i.e.~quantities that are integrated over the transverse
momentum of the Higgs boson.  Consider, however, the same integrated cross
section up to a certain transverse momentum cut $\pt^{\sss \rm cut}$.
Assuming that the Sudakov form factor provided by the \NLOPS{} implementation
is next-to-leading log~(NLL) accurate, missing next-to-next-to-leading
log~(NNLL) corrections to this quantity (corresponding to a factor $\sim
\exp[ \as^2 L] \sim 1+\as^2 L+...$, where $L=\log(\MH/\pt^{\sss \rm cut})$,
and $\MH$ is the Higgs boson mass) reduce the accuracy of the distribution,
depending upon the size of $\as L$.  If we assume $\as L^2\sim 1$, as in the
vicinity of the Sudakov region, these neglected NNLL terms are of relative
order $\as^{1.5}$ with respect to the Born term. Thus, the NLO quality (that
requires neglected terms to be of relative order $\as^2$ or higher) is
lost. Thus, although the inclusive distributions are NLO accurate, the cut
cross section may well not be if the cut is too low. Of course, if the cut is
high enough the NLO precision is restored, since \NLOPS{} simulations
describe the high transverse-momentum tail with tree-level accuracy and the
fully integrated spectrum at the NLO level.

Turning now to look at the $\HJ{}$ generator from the same perspective, if we
consider the cross section with a transverse momentum cut $\pt>\pt^{\sss \rm
  cut}$, we know that large logarithmic corrections of the form $\as L^2$
relative to the Born term arise in the cross section, hence, if $\as L^2 \sim
1$, the predictions completely fail.  Some of the new NLO merging techniques
attempt to resum these logarithmic corrections to all orders, in general
though this resummation is, with the notable exception of
ref.~\cite{Alioli:2012fc}, limited to NLL contributions and with the neglect
of the NNLL terms, as in the \H{} case, leading to errors at the level of
$\sim \as^{1.5}$ terms relative to the leading-order cross section.

In order to avoid this problem, the authors of ref.~\cite{Frederix:2012ps},
conservatively, use merging scales which are not too low. The authors of
ref.~\cite{Lonnblad:2012ix} proposes to tackle the problem by forcing
unitarity on the approach, using suitable subtractions in order to restore
NLO accuracy. In ref.~\cite{Alioli:2012fc} the accuracy of the resummation is
improved in order to reach the required precision for matching.

In the present work we study the possibility of building a NLO generator that
does not lose NLO accuracy when the radiated parton is integrated over. In
other words, we want to achieve the goals of merging without actually doing
any merging at all.

In a recent paper~\cite{Hamilton:2012np}, some of us have proposed a
prescription (\MINLO{}: Multi-scale Improved NLO) for the choice of
renormalization and factorization scales and the inclusion of Sudakov form
factors in NLO calculations, such that the higher-multiplicity calculations
seem to merge well with the lower-multiplicity ones. The numerical results of
ref.~\cite{Hamilton:2012np} are quite striking, in that they seem to suggest
that with this method, for example, the \HJJ{} simulation yields a good
description also for observables that are best computed with the \HJ{} or
\H{} programs.  In ref.~\cite{Hamilton:2012np}, however, no arguments are
given to justify this behaviour, and no attempt is made to quantify the
formal accuracy that a generator which is NLO accurate in the description of
$(n+1)$-jet observables has for those sensitive to $m$ jet topologies, with
$m \le n$.

In the present work, we address precisely these issues. We consider \BJ{}
generators, where \B{} denotes either the Higgs (\H{}), the $W$ (\W{}) or the
$Z$ boson (\Z{}) and ask the following questions:
\begin{itemize}
  \item How precise is the \BJ{} generator, when improved with the
  \MINLO{} prescription, in describing inclusive boson production
  observables?
  \item Is it possible to modify the \MINLO{} procedure in such a way
    that we achieve NLO accuracy also for these observables?
\end{itemize}
We will show that:
\begin{itemize}
  \item The inclusive boson observables are described by the \BJ{}+\MINLO{}
    programs at relative order $\as$ with respect to the Born cross
    section. However, they do not reach NLO accuracy, since they also include
    ambiguous contributions of relative order $\as^{1.5}$, rather than
    $\as^2$.
  
  \item The second question has a positive answer. In the main body of the
    paper we will clarify what needs to be done in order to gain genuine NLO
    precision for inclusive observables.  The modifications needed for
    reaching this goal are simple and we have implemented them in our current
    \MINLO{}-\POWHEGBOX{} \HJ{}, \WJ{} and \ZJ{} generators.
  \item As a final point, we notice that by reweighting the new
    \BJ{}-\MINLO{} generators in such a way that the inclusive boson rapidity
    and decay kinematics are matched to the next-to-next-to-leading
    order~(NNLO) result, one obtains a NNLO calculation matched to a parton
    shower simulation, i.e.~a \NNLOPS{} generator.
\end{itemize}
The event generators built in this way can be further enhanced, so as to
yield an improved description also for the third hardest emission, as
proposed in ref.~\cite{Alioli:2011nr}. To this end, it is enough to simply
feed the events obtained with the \BJ{}-\MINLO{} programs through the
relevant \BJJ{} simulation, with the latter attaching the third hardest
parton to the events with LL+LO accuracy.

The paper is organized as follows. In Section~\ref{sec:accuracy} we discuss
the precision of the original \BJ+\MINLO{} generator and the modifications
that are necessary to attain NLO accuracy in the description of inclusive
boson distributions. In Section~\ref{sec:proof2} we give an alternative, more
detailed proof.  In Section~\ref{sec:pheno} we present results obtained with
our revised \MINLO{} approach and discuss issues related to the scale
variation as an estimator of the theoretical error on the predictions.  In
Section~\ref{sec:NNLO} we show how we can improve the generators that we have
constructed in order to achieve NNLO accuracy.  In
Section~\ref{sec:Prospects} we discuss briefly how the present findings may
be generalized to more complex processes.  Finally we present our conclusions
in Section~\ref{sec:conclu}.

\section{Accuracy of the \BJ{}+\MINLO{} generators}
\label{sec:accuracy}
\subsection{Preliminaries}
We want to determine the accuracy of the \BJ{}+\MINLO{} generators when
integrated over the boson transverse momentum at fixed rapidity. We will
illustrate explicitly the case of Higgs boson plus jet production, but our
findings hold trivially for \WJ{} and \ZJ{} as well, and for the production
of a heavy colour-neutral system accompanied by one jet.

In order to avoid confusion, we will refer to the accuracy of inclusive
results, integrated over the boson transverse momentum, as the $\left( 0
\right)$ accuracy, and will refer instead to the accuracy in the $H + j$
inclusive cross section as the $\left( 1 \right)$ accuracy. Thus, for
example, by LO$^{\left( 0 \right)}$ and NLO$^{\left( 0 \right)}$ we mean
leading (i.e.~$\as^2$) and next-to-leading (i.e.~$\as^3$) order accuracy in
the Higgs boson inclusive rapidity distribution, and with LO$^{\left( 1
  \right)}$ and NLO$^{\left( 1 \right)}$ we mean leading (i.e.~$\as^3$) and
next-to-leading (i.e.~$\as^4$) order accuracy in the Higgs boson plus one jet
distributions.

We focus upon the \HJ{} generator with the \MINLO{} prescription. As
illustrated in~\cite{Hamilton:2012np}, this is obtained by modifying the
\POWHEG{} $\bar{B}$ function with the inclusion of the Sudakov form factor
and with the use of appropriate scales for the couplings, according to the
formula
\begin{equation}
\label{eq:sudagen}
\bar{B}=\as^2 \left( \MH^2 \right) \as \left( \qT^2 \right) \Delta_g^2 \left( \MH ,
  \qT \right) \left[ B \left( 1 - 2 \Delta_g^{\left( 1 \right)} \left(
    \MH, \qT \right) \right) + V + \int d \Phi_{\tmop{rad}} \,
    R\right]\,, 
\end{equation}
where we have stripped away three powers of $\as$ from the Born ($B$), the
virtual ($V$) and the real ($R$) contribution, factorizing them in front of
eq.~(\ref{eq:sudagen}). We will comment later on the scale at which the
remaining power of $\as$ in $R$, $V$ and $\Delta_g^{\left( 1 \right)}$ is
evaluated.

$\Delta_g$ is the gluon Sudakov form factor
\begin{equation}
\label{eq:gluon_sudakov}
  \Delta_g \left( Q, \qT \right) = \exp \left\{ - \int_{\qT^2}^{Q^2}
  \frac{d q^2}{q^2} \left[ A \left( \as \left( q^2 \right) \right) \log
    \frac{Q^2}{q^2} + B \left( \as \left( q^2 \right) \right) \right]
  \right\},
\end{equation}
and
\begin{equation}
\label{eq:deltag1}
  \Delta_g \left( Q, \qT \right) = 1 + \Delta_g^{\left( 1 \right)}
  \left( Q, \qT \right) + \mathcal{O} \left( \as^2 \right)
\end{equation}
is the expansion of $\Delta_g$ in powers of $\as$.  In the previous
equations, $\qT$ stands for the Higgs boson transverse momentum in the
underlying Born kinematics, and $Q^2$ its virtuality.

The functions $A$ and $B$ have a perturbative expansion in terms of constant
coefficients
\begin{equation}
\label{eq:A_and_B}
A \left( \as \right) = \sum_{i = 1}^{\infty} A_i \, \as^i, \hspace{2em} B \left(
\as \right) = \sum_{i = 1}^{\infty} B_i \,\as^i \,.
\end{equation}
In \MiNLO{}, only the coefficients $A_1$, $A_2$ and $B_1$ are used.
They are known for the
quark~\cite{Kodaira:1981nh} and for the gluon~\cite{Catani:1988vd}
and are given by
\begin{eqnarray}
&& A_1^{q} = \frac{1}{2\pi}\CF, \qquad  A_2^{q} =\frac{1}{4\pi^2}\CF\,K, \qquad 
B_1^{q} = -\frac{3}{4\pi}\CF\,,\label{eq:AiBiq}\\
&& A_1^{g} = \frac{1}{2\pi}\CA, \qquad  A_2^{g} = \frac{1}{4\pi^2}\CA\,K, \qquad 
B_1^{g} = - b_0\,\label{eq:AiBig},
\end{eqnarray}
where
\begin{eqnarray}
b_0&=&\frac{11\CA-2 n_f}{12 \pi}\,,\\
K & = & \left(\frac{67}{18}-\frac{\pi^2}{6}\right)\CA-\frac{5}{9}n_f\,.
\end{eqnarray}
The $\ord{\as}$ expansion of the Sudakov form factor in
eq.~(\ref{eq:deltag1}) is given by
\begin{equation}
 \Delta_g^{\left( 1 \right)}\left( Q, \qT \right)= \as \left[ 
-\frac{1}{2} A_1 \log^2\frac{\qt^2}{Q^2} + B_1 \log\frac{\qt^2}{Q^2} \right]\,.
\end{equation}
In ref.~\cite{Hamilton:2012np}, we give a particular prescription for the
choice of the renormalization scale in the power of $\as$ accompanying $V$,
$R$ and $\Delta_g^{\left( 1 \right)}$, and for the explicit renormalization
and factorization scales present in $V$ and $R$, that we do not need to
specify at this moment. Now we will assume, at variance
with~\cite{Hamilton:2012np} (for reasons that will become clear later), that
the renormalization scale for the power of $\as$ accompanying $V$, $R$ and
$\Delta_g^{\left( 1 \right)}$ is $\qT$.

Our argument is based upon the following considerations:
\begin{enumerate}
  \item Equation~(\ref{eq:sudagen}) yields the exact NLO$^{\left( 1 \right)}$
    cross section if expanded in terms of the strong coupling (two powers of
    them evaluated at the scale $\MH$, and the remaining ones at the scale
    $\qT$), and of the parton densities evaluated at the scale $\qT$. Thus,
    when integrated in the whole phase space, keeping fixed only the Higgs
    boson $\qT$ and rapidity, and expanded up to order $\as^4$,
    eq.~(\ref{eq:sudagen}), must include, for small $\qT$, all the singular
    parts of the NLO$^{\left( 1 \right)}$ result for the $H + j$ cross
    section.
  
  \item The singular parts of the $H + j$ cross section at NLO$^{\left( 1
    \right)}$ can also be obtained using the NNLL resummation formula for the
    Higgs boson transverse momentum.
 
  \item The NNLL formula, plus the regular part of the LO$^{\left( 1
    \right)}$ $H + j$ cross section, when integrated over all values of
    $\qT$, is NLO$^{\left( 0 \right)}$ accurate. This is due to the fact
    that, up to order $\as^3 \left( \MH^2 \right)$, the NNLL formula includes
    terms of the following form:
  \begin{equation}
    \delta \left( \qT^2 \right), \hspace{2em} \left( \frac{\log \qT^2}{\qT^2}
    \right)_+, \hspace{2em} \left( \frac{1}{\qT^2} \right)_+\,,
      \end{equation}
    and thus, when adding to it the regular terms, we achieve NLO$^{\left( 0 \right)}$
    accuracy.

  \item By comparing eq.~(\ref{eq:sudagen}) and the NNLL formula, we can thus
    see what is missing in eq.~(\ref{eq:sudagen}) in order to achieve
    NLO$^{\left( 0 \right)}$ accuracy.
\end{enumerate}
Before we continue, however, we should remember that, in the \MINLO{}
formula, $\qT$ stands for the Higgs boson transverse momentum in the
underlying Born kinematics, rather than the true Higgs boson transverse
momentum. We ignore this difference, assuming that a similar formula also
holds for the underlying Born transverse momentum. Later on we will modify
the \MINLO{} prescription to achieve full consistency.

We adopt the following NNLL formula for the Higgs boson transverse-momentum
distribution at fixed rapidity
\begin{eqnarray}
  \frac{\mathd \sigma}{\mathd y \mathd \qT^2} &=& 
  \sigma_0 \, \frac{\mathd}{\mathd \qT^2} \Big\{ 
\left[ C_{g a} \otimes f_{a / A} \right] \left( x_A, \qT \right) \, \times\,
\left[ C_{g b} \otimes f_{b / B} \right] \left( x_B, \qT\right)  
\nonumber\\
&&  \times \exp {\mathcal{S}} \left( Q, \qT \right) \Big\} + R_f,
  \label{eq:dsigdpt}
\end{eqnarray}
where $f_{a / A} $ and $f_{b / B} $ are the parton distribution functions for
partons $a$ and $b$ in hadrons $A$ and $B$, respectively, $x_A$ and
$x_B$ denoting the partons' momentum fractions
at the point of annihilation ($y=\frac{1}{2}\log\frac{x_A}{x_B}$). The
convolution operator, $\otimes$, and Sudakov exponent, $\mathcal{S}$,
are defined 
\begin{equation}
\left( f \otimes g \right) \left( x \right) \equiv \int^1_x  \frac{d
   z}{z} f \left( z \right) g \left( \frac{x}{z} \right),\qquad \quad
\exp {\mathcal{S}} \left( Q, \qT \right) \equiv
   \Delta^2_g \left( Q, \qT \right) , \label{eq:calSdef}
\end{equation}
with the coefficient functions $C_{ij}$ having the following perturbative
expansion
\begin{equation}
C_{ij}\(\as,z\) = \delta_{ij} \delta(1-z) + \sum_{n=1}^{\infty} \as^n C_{ij}^{(n)}(z) \,.
\end{equation}
Lastly, we have used  $R_f$ to label the non-singular part of the cross section. The renormalization
and factorization scales in eq.~(\ref{eq:dsigdpt}) are set to $\qT$, as
indicated explicitly in the argument of the convolution.

Transverse-momentum resummation is usually performed in impact-parameter
space. A formulation of resummation in $\qT$ space for vector-boson
production of the form~(\ref{eq:dsigdpt}) is given in
ref.~\cite{Ellis:1997ii}. It is however integrated in rapidity. Furthermore
it is not of NNLL, and not even of NLL order in the usual
sense~\cite{Frixione:1998dw}.  In Appendix~\ref{app:qtresum} we show that it
is however possible to derive a complete NNLL resummation formula in $\qT$
space starting from the $b$-space formula. It will become clear in the
following that the difference of this formula from the one of
ref.~\cite{Ellis:1997ii} is in fact irrelevant for our proof.

In the following, together with the explicit values of $A_1$, $A_2$ and
$B_1$, we need the expression for the $B_2$ coefficient too.  For Higgs boson
production, it was computed in ref.~\cite{deFlorian:2000pr}. Including the
modification needed to go from the impact-parameter space to the
transverse-momentum expression (see ref.~\cite{Ellis:1997ii} and also
Appendix~\ref{app:qtresum}) it assumes the form\footnote{Notice that, because
  of the expression of the Sudakov form factor in eq.~(\ref{eq:calSdef}), our
  $A$ and $B$ coefficients are divided by 2 with respect to those of
  refs.~\cite{deFlorian:2000pr, Davies:1984hs, Davies:1984sp}.}
\begin{eqnarray}
\label{eq:B2g}
B_2^{g({\rm H})}&=&\frac{1}{2\pi^2} \left[
\( \frac{23}{24}+\frac{11}{18}\pi^2-\frac{3}{2}\zeta_3\) \CA^2 + \frac{1}{2}
\CF n_f - \(\frac{1}{12}+\frac{\pi^2}{9}\) \CA  n_f -\frac{11}{8} \CA\CF\right]
\nonumber \\
&&\mbox{} + 4\zeta_3 (A_1^g)^2\,,
\end{eqnarray}
while for Drell-Yan production it was computed
in~\cite{Davies:1984hs,Davies:1984sp} and is given by
\begin{eqnarray}
\label{eq:B2q}
B_2^{q({\rm DY})}&=& \frac{1}{2\pi^2} \left[
\( \frac{\pi^2}{4}-\frac{3}{16}-3\zeta_3\) \CF^2 + \(\frac{11}{36} \pi^2
-\frac{193}{48} +\frac{3}{2}\zeta_3\)\CF\CA  +
\(\frac{17}{24}-\frac{\pi^2}{18}\)\CF n_f \right]
\nonumber \\
&&\mbox{}+ 4\zeta_3 (A_1^q)^2\,.
\end{eqnarray}

\subsection{The NLO$^{\left( 0 \right)}$ accuracy and \BJ+\MINLO}
We now examine in more details under which conditions eq.~(\ref{eq:dsigdpt})
achieves NLO$^{\left( 0 \right)}$ accuracy. By integrating it in $\mathd
\qT^2$, we get
\begin{equation}
  \frac{\mathd \sigma}{\mathd y} = \sigma_0 \,
  \left[ C_{g a}\otimes f_{a / A} \right] \left( x_A, Q \right) \,\times\,
  \left[ C_{g b}\otimes f_{b / B} \right] \left( x_B, Q \right) 
+ \int \mathd \qT^2 \,R_f\,,
  \label{eq:sigtot}
\end{equation}
where we have neglected the lower bound of the integration of the first term,
since, at small $\qT$, the integrand is strongly suppressed by the Sudakov
exponent, $\exp {\mathcal{S}} \left( Q, \qT \right)$. We thus see that, in
order to reach NLO$^{\left( 0 \right)}$ accuracy, the $C_{g i}$ and $C_{g j}$
functions should be accurate at order $\as$ (i.e.~they should include the
$C^{\left( 1 \right)}_{ij}$ term) and $R_f$ should be LO$^{\left( 1 \right)}$
accurate.

Notice that the form of eq.~(\ref{eq:dsigdpt}) (i.e.~the fact that it is
given as a total derivative) is such that the NLO$^{\left( 0 \right)}$
accuracy is maintained by construction, independently of the particular form
of the Sudakov form factor, as long as one includes the $C^{\left( 1
  \right)}_{i j}$ terms.

We now want to show that even if we take the derivative in
eq.~(\ref{eq:dsigdpt}), and discard terms of higher order in $\as$, the
NLO$^{\left( 0 \right)}$ accuracy is maintained. We consider
eq.~(\ref{eq:dsigdpt}) after the derivative is taken, for $\qT$. In order to
maintain its relative NLO$^{\left( 1 \right)}$ accuracy, we need to include
the $C^{\left( 1 \right)}_{i j}$ terms, the $A_1$ and $A_2$ terms, and the
$B_1$ and $B_2$ terms of eq.~(\ref{eq:A_and_B}).  After the derivative is
taken, we get terms of the following form
\begin{equation}
\label{eq:orders}
  \sigma_0 \,\frac{1}{\qT^2} \left[ \as,\,
  \as^2,\, \as^3,\, \as^4,\, \as L,\, \as^2 L,\, \as^3 L,\,
  \as^4 L \right] \exp {\mathcal{S}} \left( Q, \qT
  \right), 
\end{equation}
where $L=\log Q^2/\qT^2$ and the $\as L$ and $\as^2 L$ terms arise from the
$A$ term in the derivative of the Sudakov exponent. Some terms of order $\as$
and $\as^2$ also arise in this way, from the $B$ term. Others arise from the
derivative of the parton distribution functions (pdfs). Terms of order
$\as^4$ arise, for example, from two $C^{\left( 1 \right)}_{i j}$ terms
together with the $B_2$ term in the derivative of the Sudakov exponent. All
powers of $\as$ in the square bracket of eq.~(\ref{eq:orders}), and the
relevant pdfs, are evaluated at $\qT$.

If we do not drop any higher-order terms from eq.~(\ref{eq:orders}), its
integral is still NLO$^{\left( 0 \right)}$ accurate. This is, of course, the
case, since the formula can be written back as an exact derivative, its
integral is given by eq.~(\ref{eq:sigtot}), and the $\text{$C^{\left( 1
    \right)}_{i j}$}$ terms are included. We now want to show that even if we
drop the higher-order terms, the NLO$^{\left( 0 \right)}$ accuracy is not
spoiled.

We estimate the size of each contribution using the formula
\begin{equation}
  \label{eq:countlogs}
  \int^{Q^2}_{\Lambda^2} \frac{\mathd \qt^2}{\qt^2} \log^m \frac{Q^2}{\qt^2} \,\as^n \left( \qt^2
  \right) \,\exp {\mathcal{S}} \left( Q, \qt \right) \approx \left[
  \as \left( Q^2 \right) \right]^{n - \frac{m + 1}{2}} \,.
\end{equation}
This formula is a consequence of the fact that the dominant Sudakov
singularities carry two logarithms for each power of $\as$, and thus each
logarithm counts as $1/\sqrt{\as}$. We give however a detailed derivation of
this formula, including the effect of the running coupling, in
Appendix~\ref{sec:integrals}.

Among all terms of order higher than $\as^2$ in the square bracket of
eq.~(\ref{eq:orders}), the dominant one is $\as^3 L$. This term gives a
contribution of relative order $\left[ \as \left( Q^2 \right) \right]^{3 -
  \frac{2}{2}} = \as^2 \left( Q^2 \right)$. Thus, all terms of order $\as^3$
and higher can be dropped without spoiling the NLO$^{\left( 0 \right)}$
accuracy. Dropping these terms, we get essentially the full singular part of
the \MINLO{} \HJ{} formula, except that the original \MINLO{} formula does
not have the $B_2$ term in $\mathcal{S}$. In fact, if we expand the Sudakov
factor up to ${\cal O}(\as)$ and combine it with the content of the squared
parenthesis, we get the full singular part of the \HJ{} cross section. Since
also the \MINLO{} \HJ{} formula has the same property, the two must agree.
We also remark that the choice of the scale in the power of $\as$ entering
$V$, $\Delta^{\left( 1 \right)}$ and $R$ (that we have taken equal to $\qT$)
is essential for our argument to work.  For example, if we choose instead a
scale equal to $Q$, the largest difference arises in the terms of order
$\as^2 L$ in eq.~(\ref{eq:orders}), where they would give an $\mathcal{O}
\left( \as^3 L^2 \right)$ variation. This yields a contribution of order
$\mathcal{O} \left( \as^{1.5} \left( Q^2 \right) \right)$ upon
integration. On the other hand, we will show in the following this kind of
contributions are of the same size as the effect of the $B_2$ term. Thus, if
the latter is not included, this scale choice remains ambiguous.

We now investigate what is the loss of precision due to the lack of the $B_2$
term in the \MINLO{} formula. In order to do this, we drop the $B_2$ term in
the Sudakov exponent in eq.~(\ref{eq:dsigdpt}). The resulting formula still
satisfies eq.~(\ref{eq:sigtot}), and thus is NLO$^{\left( 0 \right)}$
accurate. When taking the derivative, however, we will get all the terms in
the square bracket of eq.~(\ref{eq:orders}), except for the term
\begin{equation}\label{eq:b2term}
  \sigma_0 \, \frac{1}{\qT^2} \, \as^2(\qt^2) \,B_2\, \exp {\mathcal{S}} \left( Q, \qT \right) .
\end{equation}
This term is instead present in the \MINLO{} result, that agrees with
eq.~(\ref{eq:orders}) up to the terms of order $\as^2$ in the square bracket,
and differs from it only by subleading terms in the Sudakov exponent.  We
conclude that, if we take away the contribution of eq.~(\ref{eq:b2term}) from
the \MINLO{} result, we recover the NLO$^{\left( 0 \right)}$ accuracy. Thus,
the \MINLO{} formula, as is, violates the NLO$^{\left( 0 \right)}$ accuracy
by this term, which yields, according to eq.~(\ref{eq:countlogs}), a
contribution of relative order $\mathcal{O} \left( \as^{1.5} \left( Q^2
\right) \right)$ upon integration.

\subsection{Summary}
Summarizing our findings:
\begin{itemize}
\item the original \BJ{}+\MINLO{} generator of~\cite{Hamilton:2012np}, is
  less than NLO$^{\left( 0 \right)}$ accurate, in that it includes incorrect
  terms of relative order \ $\mathcal{O} \left( \as^{1.5} \left( Q^2 \right)
  \right)$.
  
  \item In order to achieve NLO$^{\left( 0 \right)}$ accuracy for the
    \BJ+\MINLO{} generator, we must
  \begin{itemize}
    \item include the $B_2$ term in the Sudakov form factor of
      eq.~(\ref{eq:gluon_sudakov});
    
    \item take the scale of the power of $\as$ entering $V$,
    $\Delta^{\left( 1 \right)}$ and $R$ equal to $\qT$.
  \end{itemize}
\end{itemize}
In this section we have not considered the possibility of varying the
factorization and renormalization scales in the \MiNLO{} improved $\tilde{B}$
functions. It is indeed possible to vary them in such a way that the
NLO$^{\left( 0 \right)}$ accuracy for the \BJ+\MINLO{} generator is
maintained. In Appendix~\ref{app:scalevar} we show in detail how this can be
done.

\section{Secondary proof of NLO accuracy}
\label{sec:proof2}
In this section we describe an independent derivation, corroborating that of
Sect.~\ref{sec:accuracy}, that \MiNLO{} heavy boson plus jet production
computations, with minor modifications, yield predictions NLO accurate in the
description of both inclusive and jet-associated observables.  The
alterations to the original \MiNLO{} algorithm required here are the same as
those discussed previously in Sect.~\ref{sec:accuracy}. Principally this
amounts to using the transverse momentum of the heavy boson, as opposed to
the scale obtained on clustering the Born kinematics with the $\kT$-jet
algorithm, as an input to the pdfs, strong coupling and Sudakov form factor,
together with the inclusion of the NNLL $B_2$ coefficient in the latter. The
analysis essentially comprises of four steps:
\begin{enumerate}
\item Using existing expressions for their singular parts, we write the NLO
  cross section for heavy boson plus jet production, differential in the
  boson's rapidity, $y$, and transverse momentum, $\pTKH $, as the sum of a
  $\pTKH \rightarrow 0$ divergent part, $\mathd \sigma_{{\scriptscriptstyle
      \mathcal{S}}}$, and a finite remainder, $\mathd
  \sigma_{{\scriptscriptstyle \mathcal{F}}}$, with scales set according to
  the slightly revised \MiNLO{} conventions.

\item The \MiNLO{} Sudakov form factor and coupling constant weights are
  applied to the resulting expression and the related subtraction terms,
  necessary to maintain NLO accuracy in the presence of these weights, are
  inserted.

\item By neglecting terms which, on integration over $\pTKH $, give rise to
  unenhanced terms $\mathcal{O}\left(\as^{2}\right)$ with respect to the
  lowest order heavy boson production process, we are able to cast the
  singular part of the \MiNLO{} cross section in the form of a total
  derivative.

\item The $\pTKH $ integration of the singular cross section may be performed
  at NLO accuracy and combined with the leading part of the remainder cross
  section, $\mathd \sigma_{{\scriptscriptstyle \mathcal{F}}}$, whereupon the
  sum can be identified as the NLO differential cross section.
\end{enumerate}
In the following we give details regarding these steps, pointing out any
deviations with respect to the original \MiNLO{} algorithm as they enter.
The mathematical description of the algorithm, as given here, reflects,
precisely, that of the implementation whose results we present in
Sect.~\ref{sec:pheno}. Lastly, we further stress that although the formulae
we use explicitly refer only to the heavy bosons' rapidity distributions, the
derivations hold, with a simple modification, also for the case of
distributions involving the heavy-boson decay products. This generalization
is discussed in Sect.~\ref{sub:extension-to-vector-boson-decays}.

\subsection{NLO $\pTKH $ spectra with \MiNLO{} scale choices}
\label{sub:initial-scale-settings}
To begin proving the NLO accuracy of the \MiNLO{} predictions for inclusive
observables, we require an expression for the cross section differential in
the transverse momentum and rapidity of the heavy boson, employing the
renormalization and factorization scales set out in
ref.~\cite{Hamilton:2012np}. Contrary to the original \MiNLO{} prescription,
however, we now insist that the transverse momentum of the massive boson be
used as the factorization scale, moreover, we now also wish that this scale,
$\pTKH $, be used as the argument in evaluating the extra $\as$ factor
accompanying the NLO corrections $\left(\as^{\mathrm{{\scriptscriptstyle
      NLO}}}\right)$.

The asymptotic $\pTKH \rightarrow0$ limit of $W$ and $Z$ transverse momentum
spectra was given in ref.~\cite{Arnold:1990yk} to next-to-leading order
accuracy. The NLO Higgs boson transverse momentum spectrum has also been
computed in ref.~\cite{Glosser:2002gm}.  In the latter work the authors
derived the limit directly from their NLO computation and found agreement
with the expression obtained by taking the analogous limit of the $b$-space
resummation formula of ref.~\cite{Collins:1984kg}. For $W$, $Z$ and Higgs
boson production processes the results are quoted for arbitrary
renormalization~($\mur$) and factorization scales~($\muf$).

Using these results one can directly write down NLO expressions for the small
transverse momentum limit of the heavy-boson $\pTKH $ spectra, by simply
replacing all instances of $\mur$ and $\muf$ (explicit and implicit) by $\KRA
Q$ and $\KFA\pTKH $, respectively, in the asymptotic expressions of
refs.~\cite{Arnold:1990yk,Glosser:2002gm}.  $\KFA$ and $\KRA$ simply denote
constant rescaling factors used to perform scale uncertainty estimates, thus
they are nominally set equal to one, nevertheless, we retain them explicitly
in order to better clarify the nature of such variations.  Having made the
aforesaid replacements, one can trivially substitute the additional power of
$\as$ in the NLO terms by $\as^{\mathrm{{\scriptscriptstyle NLO}}}
=\as(\KRA^2 Q^2) $ simply, with no further changes of any kind.

We thus write the $\pTKH $ spectrum of a heavy boson resulting from the
annihilation of partons $i$ and $j$, from beam particles $A$ and $B$
respectively, as the sum of a $\pTKH \rightarrow0$ singular contribution,
$\mathd \sigma_{{\scriptscriptstyle \mathcal{S}}}$, and a finite regular
remainder term, $\mathd \sigma_{{\scriptscriptstyle \mathcal{F}}}$,
\begin{equation}
\label{eq:nlo-hj-muR-KRmH-muF-KFpT}
\frac{\mathd \sigma}{\mathd\pTKH ^{2}\mathd y}= \frac{\mathd
  \sigma_{{\scriptscriptstyle \mathcal{S}}}}{\mathd\pTKH ^{2}\mathd y}+
\frac{\mathd \sigma_{{\scriptscriptstyle \mathcal{F}}}}{\mathd\pTKH
  ^{2}\mathd y}\,,
\end{equation}
where the singular part takes, precisely, the form 
\begin{equation}
\label{eq:Glosser-Schmidt-pT-singular-formula}
\frac{\mathd \sigma_{{\scriptscriptstyle \mathcal{S}}}}{\mathd\pTKH
  ^{2}\mathd y} = \frac{N}{\pTKH ^{2}}\,\left(\frac{\as}{2\pi}\right)^{n}\,
\sum_{r=1}^{2}\,\sum_{s=0}^{2r-1}\,\left(\frac{\as}{2\pi}\right)^{r}\,{}_{r}D_{s}\,
L^{s}\,,
\end{equation}
with $L=\log\left(Q^2/\pTKH ^{2}\right)$. 
The $_{r}D_{s}$ coefficients are directly obtained from the $_{m}C_{n}$
coefficients given in Appendix~A.2 of ref.~\cite{Arnold:1990yk} and
Appendix~C of ref.~\cite{Glosser:2002gm}, following the replacements
described above. Since the resulting expressions are lengthy but trivial to
derive we refrain from quoting them explicitly here. Lastly, in the
normalization of eq.~(\ref{eq:Glosser-Schmidt-pT-singular-formula}) the
factor of $\as^{n}$ accounts of there being $n$ powers of the strong coupling
associated with the leading order production process ($n=0$ for $W$ and $Z$
production, $n=2$ for Higgs boson production) while $N$ comprises constant
factors such that the product of $N\left(\as/2\pi\right)^{n}$ with the pdfs
gives the LO cross section for $H/W/Z$ production in the $i,\,j$ channel,
differential in $y$.

Equation~(\ref{eq:nlo-hj-muR-KRmH-muF-KFpT}) describes, precisely, the cross
section returned by the \MiNLO{} programs prior to the introduction of
Sudakov form factor and coupling constant reweightings.  In particular, since
the cross section we write here is intended to refer to that coded in the
programs, there are no unknown $\mathcal{O}\left(\as^{n+3}\right)$ terms or
beyond omitted anywhere, throughout it contains explicitly only contributions
proportional to $\as^{n+1}$ and $\as^{n+2}$, with all strong coupling and
pdfs factors utilizing the scale settings already elaborated on.

\subsection{Differential cross section in the \MiNLO{} programs}
\label{sub:MiNLO-exact-xsecs}
Following the \MiNLO{} procedure we must now multiply the NLO cross section
by a Sudakov form factor and, to maintain NLO accuracy, simultaneously
subtract a term corresponding to the $\mathcal{O}\left(\as^{n+2}\right)$
contribution implicit in the product of the leading order terms with the form
factor. In so doing, as was with the choice of factorisation scale, here we
wish to change the lower scale entering the Sudakov form factor, from that
obtained on the first clustering of the underlying Born kinematics with the
$\kT$-jet algorithm, to simply $\pTKH $.

In our programs we have used the following Sudakov form factor, describing
the evolution of a quark / gluon line from scale $Q$ to $\pTKH $, generalised
to account for a renormalization scale rescaling factor $\KRA$
\begin{equation}
\log\Delta\left(Q,\KRA \pTKH\right)=-\,\int_{\KRA^2\pTKH
  ^{2}}^{Q^{2}}\,\frac{\mathd
  q^{2}}{q^{2}}\,\left[\tilde{A}\left(\as\left(q^2\right)\right)\log\frac{Q^2}{q^{2}}
+ \tilde{B}\left(\as\left(q^2\right)\right)\right]\,,
\end{equation}
wherein 
\begin{equation}
\label{eq:minlo-A-B-tilde-defn}
\tilde{A}\left(\as\right)=\sum_{i=1}^{\infty}\as^{i}\,\tilde{A}_{i}\,, 
\qquad {\rm and} \qquad
\tilde{B}\left(\as\right)=\sum_{i=1}^{\infty}\as^{i}\,\tilde{B}_{i}\,,
\end{equation}
with the $\tilde{A}_{i}$ and $\tilde{B}_{i}$ coefficients related to the
$A_{i}$ and $B_{i}$ of Sect.~\ref{sec:accuracy} as follows
\begin{eqnarray}
\tilde{A}_{1} & = & A_{1}\\
\tilde{A}_{2} & = & A_{2}+A_{1}\, b_{0}\,\log\, \KRA^2\\
\tilde{B}_{1} & = & B_{1}+A_{1}\,\log\, \KRA^2\\
\label{eq:B2tilde}
\tilde{B}_{2} & = & B_{2}^{{\scriptscriptstyle \left(\mathrm{X}\right)}}
+\frac{1}{2}nb_{0}^{2}\,\log\,
\KRA^2 + A_{2}\,\log\, \KRA^2+\frac{1}{2}\, A_{1}\, b_{0}\,\log^{2} \KRA^2\,.
\end{eqnarray}
Here, in the definition of $\tilde{B}_{2}$, we have used
$B_{2}^{{\scriptscriptstyle \left(\mathrm{X}\right)}}$ to denote either
$B_2^{g({\rm H})}$ or $B_2^{q({\rm DY})}$, as defined in eqs.~(\ref{eq:B2g})
and~(\ref{eq:B2q}) respectively.  The derivation of the scale dependence of
these coefficients is given in Appendix~\ref{app:scalevar}.

For the case of NLO Higgs or vector boson plus jet production processes, the
\MiNLO{} algorithm will multiply the full cross section by the product of two
such Sudakov form factors. Introducing the abbreviation, $\tilde{L}=L-\log
\KRA^2$, the first order expansion of this aggregate form factor is given by
\begin{eqnarray}
\label{eq:Sudakov-expanded-in-aS}
\Delta^2\!\left(Q,\KRA \pTKH \right) & = & 1-\as\left[\tilde{A}_{1}\,
  \tilde{L}^{2}+2\,\tilde{B}_{1}\,\tilde{L}\right]+\mathcal{O}
\left(\as^{2}\tilde{L}^{4}\right)\,.
\end{eqnarray}
To compensate the spurious $\mathcal{O}\left(\as^{n+2}\right)$ contribution
which this will generate, when it multiplies the leading order part of the
heavy boson plus jet cross section (distinguished by the subscript
$\mathrm{LO}$), a corresponding counterterm is added.

In addition to the Sudakov form factor, the \MiNLO{} procedure, for Higgs,
$W$ and $Z$ plus jet processes, prescribes that the whole cross section be
multiplied by an overall factor comprised of a ratio of coupling constants
\begin{equation}
\as\left(\KRA^2 \pTKH^2 \right)/\as\left(\KRA^2 Q^2\right)=1+\as\, b_{0}\,
L+\mathcal{O}\left(\as^{2}L^{2}\right)\,.\label{eq:alphaS-ratio-expansion}
\end{equation}
As for the factorisation scale and the low scale for the Sudakov form factor,
here we have revised the original \MiNLO{} algorithm by taking the scale in
$\as$ in the numerator to be $\KRA \pTKH $.  To apply this overall weight
factor and maintain the integrity of the NLO cross section, a counterterm is
introduced in the \MiNLO{} cross section, balancing the erroneous
contribution which it elicits at $\mathcal{O}\left(\as^{n+2}\right)$.

Taking into account the Sudakov form factor, the coupling constant ratios,
and their respective counterterms, the \MiNLO{} cross section, $\mathd
\sigma_{{\scriptscriptstyle \mathcal{M}}}$, reads as follows
\begin{eqnarray}
\label{eq:nlo-hj-minlo-initial}
\frac{\mathd \sigma_{{\scriptscriptstyle \mathcal{M}}}}{\mathd\pTKH
  ^{2}\mathd y} & = & \Delta^2\!\left(Q,\KRA \pTKH
\right)\,\frac{\as\left(\KRA^2 \pTKH^2 \right)}{\as\left(\KRA^2 Q^2\right)}
\nonumber\\
 & & {}\times \left[\frac{\mathd \sigma}{\mathd\pTKH ^{2}\mathd y} +
  \left.\frac{\mathd \sigma}{\mathd\pTKH ^{2}\mathd
    y}\right|_{{\scriptscriptstyle \mathrm{LO}}}\,\as^{{\scriptscriptstyle
      \mathrm{NLO}}}\left[\tilde{A}_{1}\tilde{L}^{2}+2\tilde{B}_{1}\tilde{L}-b_{0}\left(\tilde{L}+\log
    \KRA^2\right)\right]\right]\,.
\end{eqnarray}
All modifications explained in this cross section have been defined
precisely, with no ambiguities of any kind, higher order or otherwise, that
is to say, eq.~(\ref{eq:nlo-hj-minlo-initial}) reflects, exactly, the
implementation of the differential cross section in the \MiNLO{} programs for
Higgs, $W$ and $Z$ plus jet production.

Combining eqs.~(\ref{eq:nlo-hj-muR-KRmH-muF-KFpT}),
(\ref{eq:Glosser-Schmidt-pT-singular-formula})
and~(\ref{eq:nlo-hj-minlo-initial}), using only basic algebraic
manipulations, we may write the cross section from the \MiNLO{} programs in
such a way as to isolate the leading $\pTKH \rightarrow0$
behaviour. Specifically, we further develop our $\mathd
\sigma_{{\scriptscriptstyle \mathcal{M}}}$ formula as
\begin{eqnarray}
\label{eq:nlo-hj-minlo-fully-applied}
\frac{\mathd \sigma_{{\scriptscriptstyle \mathcal{M}}}}{\mathd\pTKH
  ^{2}\mathd y} & = & \frac{\mathd \sigma_{{\scriptscriptstyle
      \mathcal{MS}}}}{\mathd\pTKH ^{2}\mathd y}+\frac{\mathd
  \sigma_{{\scriptscriptstyle \mathcal{MF}}}}{\mathd\pTKH ^{2}\mathd
  y}\,,
\end{eqnarray}
with the leading $\pTKH \rightarrow0$ part given by
\begin{equation}
\label{eq:Glosser-Schmidt-pT-singular-formula-1}
\frac{\mathd \sigma_{{\scriptscriptstyle \mathcal{MS}}}}{\mathd\pTKH
  ^{2}\mathd y}=\Delta^2\!\left(Q,\KRA \pTKH \right)\,\frac{N}{\pTKH
  ^{2}}\,\left(\frac{\as}{2\pi}\right)^{n}\,\sum_{r=1}^{2}\sum_{s=0}^{1}\,\as^{r}\,{}_{r}E_{s}\,\tilde{L}^{s}\,,
\end{equation}
and subleading part given by eq.~(\ref{eq:nlo-hj-minlo-initial}) save for the
replacements $\mathd \sigma_{{\scriptscriptstyle \mathcal{M}}}\rightarrow
\mathd \sigma_{{\scriptscriptstyle \mathcal{MF}}}$ on the left- and $\mathd
\sigma\rightarrow \mathd \sigma_{{\scriptscriptstyle \mathcal{F}}}$ on the
right-hand side. In eq.~(\ref{eq:Glosser-Schmidt-pT-singular-formula-1}) $n$
powers of $\as$ are evaluated at a scale $\KRA Q$ and the remainder at $\KRA
\pTKH $.  Assuming the same shorthand as ref.~\cite{Arnold:1990yk} for the
pdfs, $f_{i}=f_{i/A}\left(x_{A},\mu_{F}\right)$, the $_{r}E_{s}$ coefficients
are found to be
\begin{eqnarray}
\label{eq:mC'n-coefficients}
_{1}E_{1} & = & 2\tilde{A}_{1}\, f_{i}\, f_{j}\,,\\
_{1}E_{0} & = & 2\tilde{B}_{1}\, f_{i}\, f_{j}+\left[P_{ik}\otimes
    f_{k}\right]\, f_{j}+\left[P_{jk}\otimes f_{k}\right]\, f_{i}\,,\nonumber
  \\
{2}E_{1} & = & \tilde{A}_{2}\, f_{i}\,
f_{j}+2\tilde{A}_{1}\,\left[C_{ik}^{\left(1\right)\prime\prime}\otimes
  f_{k}\right]\, f_{j}+\left\{ i\leftrightarrow j\right\}\,,\nonumber \\
_{2}E_{0} & = & \tilde{B}_{2}\, f_{i}\,
f_{j}+2\tilde{B}_{1}\left[C_{ik}^{\left(1\right)\prime\prime}\otimes
  f_{k}\right]\, f_{j}+\left[P_{ik}^{\left(2\right)}\otimes f_{k}\right]\,
f_{j}+b_{0}\ln\frac{\KRA^{2}}{\KFA^{2}}\,\left[P_{ik}\otimes f_{k}\right]\,
f_{j}\nonumber\\
 & &\mbox{}+ \left[C_{ik}^{\left(1\right)\prime\prime}\otimes
  f_{k}\right]\left[P_{jl}\otimes
  f_{l}\right]+\left[C_{ik}^{\left(1\right)\prime\prime}\otimes P_{kl}\otimes
  f_{l}\right]\, f_{j}-b_{0}\left[C_{ik}^{\left(1\right)\prime\prime}\otimes
  f_{k}\right]\, f_{j}+\left\{ i\leftrightarrow j\right\}\,,\nonumber
\end{eqnarray}
with $C_{ik}^{\left(1\right)\prime\prime}$ defined, exactly, as
\begin{eqnarray}
C_{ik}^{\left(1\right)\prime\prime}\left(z\right) & = & C_{ik}^{\left(1\right)}\left(z\right)+\frac{1}{2}\, n\, b_{0}\log \KRA^{2}\,\delta_{ik}\delta\left(1-z\right)-\log \KFA^{2}\, P_{ik}\left(z\right)\nonumber\\
 & &\mbox{}-\left(\frac{1}{2}\, A_{1}\,\log^{2}\KRA^{2}+B_{1}\,\log
\KRA^{2}\right)\delta_{ik}\delta\left(1-z\right)\,.
\end{eqnarray}
The $C_{ik}^{\left(1\right)}$ terms are NLO corrections to the coefficient
functions in the conventional $b$-space resummation formula, while
$P_{ik}\text{\ensuremath{\left(z\right)}}$ denotes the leading order
splitting function for parton~$k$ branching to a parton~$i$ with momentum
fraction $z$, $P_{ik}^{\left(2\right)}$ representing its relative
$\mathcal{O}\left(\as\right)$ corrections. The $C_{ik}^{\left(1\right)}$ and
$P_{ik}$ functions here are equal to those in
refs.~\cite{Arnold:1990yk,Glosser:2002gm} divided by $2\pi$, similarly, our
$P_{ik}^{\left(2\right)}$ functions are equal to those in
refs.~\cite{Arnold:1990yk,Glosser:2002gm} divided by
$\left(2\pi\right)^{2}$.%
\footnote{In simplifying the $_{r}E_{s}$ coefficients, in particular for
  collecting the $C_{ik}^{\left(1\right)\prime\prime}$ terms, the following
  trivial identities are useful:
  $f_{i}=\delta_{ik}\delta\left(1-z\right)\otimes f_{k}\,$ and $P_{ik}\otimes
  f_{k}=\delta_{ik}\delta\left(1-z\right)\otimes P_{kl}\otimes f_{l}\,.$}

In determining eq.~(\ref{eq:nlo-hj-minlo-fully-applied}) from
eq.~(\ref{eq:nlo-hj-minlo-initial}) we have at no point employed Taylor
expansions, renormalization group or DGLAP equations;
eq.~(\ref{eq:nlo-hj-minlo-fully-applied}) follows exactly from
eq.~(\ref{eq:nlo-hj-minlo-initial}) without neglect of terms at
$\mathcal{O}\left(\as^{n+3}\right)$ or beyond. Thus
eq.~(\ref{eq:nlo-hj-minlo-fully-applied}) corresponds precisely, without
ambiguities of any kind, higher order or otherwise, to the differential cross
section returned by the \MiNLO{} programs.

\subsection{Integrating the \MiNLO{} $\pTKH $ spectrum with NLO accuracy}
\label{sub:Approximations-leading-to}
In this subsection, through judicious omission of higher order contributions,
together with application of the DGLAP and coupling constant evolution
equations, we write the leading $\pTKH \rightarrow0$ part of the \MiNLO{}
cross section, $\mathd \sigma_{{\scriptscriptstyle \mathcal{MS}}}$, in the
form of a derivative with respect to $\pTKH $.  To this end we use the
following result derived from the integral in Appendix~\ref{sec:integrals}:
terms in the differential cross section
$\mathcal{O}\left(\as^{n+r}L^{s}/\pTKH ^{2}\right)$, with
$r\ge\frac{1}{2}\left(s+5\right)$, multiplied by the Sudakov form factor,
yield contributions $\mathcal{O}\left(\as^{n+2}\right)$ or higher on
integration over the low $\pTKH $ domain, i.e.~such terms do not affect
inclusive observables at the NLO level, nor, for $r\ge3$, do they affect NLO
accuracy of jet-associated $H/W/Z$ production observables.

Neglecting contributions not greater than $\mathcal{O}\left(\as^{n+3}L/\pTKH
^{2}\right)$ times the Sudakov form factor, one can replace the sum of terms
proportional to $\tilde{A}_{i}$ and $\tilde{B_{i}}$ in $\mathd
\sigma_{{\scriptscriptstyle \mathcal{MS}}}$ by the $\pTKH $ derivative of the
Sudakov form factor, multiplied by a product of the coefficient functions,
$C_{ik}^{\prime\prime}$, defined
\begin{equation}
\label{eq:Cij-prime-prime-defn}
C_{ik}^{\prime\prime}\left(\as^{{\scriptscriptstyle
\mathrm{NLO}}},z\right)=\delta_{ik}\delta\left(1-z\right)+\as^{{\scriptscriptstyle
\mathrm{NLO}}}C_{ik}^{\left(1\right)\prime\prime}\left(z\right)\,.
\end{equation}
Further, dropping terms $\mathcal{O}\left(\as^{n+3}\right)$ inside the
summation of eq.~(\ref{eq:Glosser-Schmidt-pT-singular-formula-1}), one may
substitute
\begin{equation}
\as\left(\KRA^2 \pTKH^2 \right)\left(P_{ik}+\as^{{\scriptscriptstyle
      \mathrm{NLO}}}\left(P_{ik}^{\left(2\right)}+b_{0}\,
  P_{ik}\,\log\,\frac{\KRA^2}{\KFA^2 }\right)\right)\otimes
  f_{k}\,\rightarrow\, \pTKH
^{2}\,\frac{\mathd f_{i}}{\mathd\pTKH ^{2}}
\,,\label{eq:splitting-fns-replaced-by-derivative}
\end{equation}
with the implicit scale in the pdfs being $\KFA \pTKH $
as usual.%
\footnote{Among the $\mathcal{O}\left(\as^{n+3}\right)$ terms forgone in this
  replacement are contributions proportional to
  $\as^{n+3}\log^{2}\KRA^2/\KFA^2 $, which we neglect in the understanding
  that the ratio $\KRA /\KFA $ will be of order 1. In fact, in our
  phenomenology section~\ref{sec:pheno}, the ratio is confined to the
  interval $\left[\frac{1}{2},2\right]$.}  The same replacement without the
$\as^{{\scriptscriptstyle \mathrm{NLO}}}$ terms on the left-hand side holds,
at the same level of approximation, for the two remaining terms in
$_{2}E_{0}$ containing the leading order splitting functions. Lastly,
discarding additional terms of the same order as those omitted in
eq.~(\ref{eq:splitting-fns-replaced-by-derivative}), one may also replace the
remaining $b_{0}$ factor in the $_{2}E_{0}$ coefficient as
\begin{equation}
b_{0}\rightarrow-\frac{\pTKH ^{2}}{\as^{2}}\,\frac{\mathd \as}{\mathd\pTKH ^{2}}\,,
\end{equation}
with the scale for $\as$ in the denominator and the derivative being $\KRA
\pTKH $.

With the help of these substitutions the leading $\pTKH \rightarrow0$ part of
the \MiNLO{} cross section, neglecting terms no greater than
$\sim\Delta{}^{2}\as^{n+3}L/\pTKH ^{2}$, can finally be written as
\begin{equation}
\label{eq:final-minlo-formula}
\frac{\mathd \sigma_{{\scriptscriptstyle \mathcal{MS}}}}{\mathd\pTKH
  ^{2}\mathd
  y}=N\,\left(\frac{\as}{2\pi}\right)^{n}\,\frac{\mathd}{\mathd\pTKH
  ^{2}}\,\Big\{\Delta^2\!\left(Q,\KRA
\pTKH\right)\,\left[C_{ik}^{\prime\prime}\otimes
  f_{k}\right]\,\left[C_{jl}^{\prime\prime}\otimes f_{l}\right] \Big\}\,,
\end{equation}
where the $n$ powers of $\as$ explicit in the prefactor retain $\KRA Q$ as
their argument, with $\KRA \pTKH $ holding for the rest. Integrating
eq.~(\ref{eq:final-minlo-integrated}) over $\pTKH$ yields
\begin{eqnarray}
\label{eq:final-minlo-integrated}
\frac{\mathd \sigma_{{\scriptscriptstyle \mathcal{MS}}}}{\mathd y} & = &
N\,\left(\frac{\as}{2\pi}\right)^{n}\,\Delta^2\!\left(Q,\KRA
Q\right)\,\left[C_{ik}^{\prime\prime}\otimes
  f_{k}\right]\,\left[C_{jl}^{\prime\prime}\otimes
  f_{l}\right]\,,
\end{eqnarray}
with the coupling constants in the $C_{ik}^{\prime\prime}$ having $\KRA Q$ as
their argument and the scale in the pdfs set now, here, to $\KFA Q$. The
Sudakov form factor here contains no large logarithms since the integral
range in the exponent is limited to high values, $\left[\KRA^2 Q^2,\,
  Q^2\right]$, thus we may reliably expand
eq.~(\ref{eq:final-minlo-integrated}) in powers of $\as$ neglecting NNLO
sized contributions to give simply
\begin{eqnarray}
\frac{\mathd \sigma_{{\scriptscriptstyle \mathcal{MS}}}}{\mathd y} & = &
N\,\left(\frac{\as}{2\pi}\right)^{n}\,\left\{\,
f_{i}\,f_{j}+\as\left[C_{ik}^{\left(1\right)}\otimes
  f_{k}\right]f_{j}+\as\left[C_{jk}^{\left(1\right)}\otimes
  f_{k}\right]f_{i}\right\} \,,
\end{eqnarray}
wherein all pdf and $\as$ factors are evaluated at the common scale $Q$. The
disappearance of all factors of $A_{i}$, $B_{i}$ in this expression follows
from the cancellation of the NLO expansion of the Sudakov factor,
$\Delta^2\!\left(Q,\KRA Q\right)$, with an equal and opposite contribution in
the product of the $C_{ik}^{\prime\prime}$ coefficient functions. Similarly,
the vanishing of all $\KRA $ and $\KFA $ factors attributes to the explicit
$\log K_{{\scriptscriptstyle R/F}}^{2}$ terms being absorbed in the pdfs and
strong coupling constant factors, renormalizing the scales in these factors,
in the leading order contribution, from $K_{{\scriptscriptstyle R/F}}Q$ to
$Q$. The scales of the pdfs and coupling constants in the remaining (NLO)
terms are freely changeable at this level of accuracy.

To proceed we must explain the nature of the NLO component of the coefficient
functions, $C_{ik}^{\left(1\right)}$ . These are given, essentially as a
matter of definition, by half of the sum of the virtual and real, $\pTKH
\rightarrow0$ singular, components of the NLO cross section for heavy boson
plus jet production, with the real-radiation phase space below $\pTKH =Q$
having been integrated over. This calculation is carried out explicitly in
ref.~\cite{Kauffman:1991cx}, for the case of Higgs boson production, and in
Sect.~2 of ref.~\cite{Altarelli:1984pt} for the $q\bar{q}$ channel in vector
boson production. The coefficients obtained agree precisely with those in
refs.~\cite{Glosser:2002gm, Kauffman:1991cx}, i.e.~with those employed in the
expressions for the NLO $\pTKH $-spectra from which we started our analysis
in Sect.~\ref{sub:initial-scale-settings}.  Additionally, at the end of
Appendix~C in ref.~\cite{Balazs:1997xd} the integral of the $W$ and $Z$
$\pTKH $ spectra up to some arbitrary $\pTKH $ is shown, wherein one can
readily identify the NLO $C_{ij}$ functions as the combined virtual and
integrated real, $\pTKH \rightarrow0$ singular, contributions to the weak
boson cross section.
\footnote{Up to an irrelevant notational difference of a factor of two, the
coefficient functions in ref.~\cite{Balazs:1997xd} are identical
to those of ref.~\cite{Kauffman:1991cx}.} 

Finally we assert that the integral of the remaining part of the $\pTKH$
spectrum, $\mathd \sigma_{{\scriptscriptstyle \mathcal{MF}}}$, is equal to
that of the original $\pTKH\rightarrow0$ regular component, $\mathd
\sigma_{{\scriptscriptstyle \mathcal{F}}}$, (with the coupling constant and
pdf scales of order $Q$), up to formally relative order $\as ^{2}$ terms.
This statement follows from the fact that the regular parts are suppressed by
a factor $\pTKH^{2}/Q^{2}$ relative to their diverging counterparts in
$\mathd \sigma_{{\scriptscriptstyle \mathcal{S}}}$.  Thus, in the region
$\pTKH\lesssim Q \as\left(Q^2\right)$ surrounding the Sudakov peak, the
regular piece is effectively a relative $\as^2\left(Q^2\right)$ contribution
and hence entirely negligible from the point of view of our NLO
analysis. Just outside this region the same principle applies, namely, that
provided the proportionate contribution made by the regular piece is small
with respect to the finite ones, the impact that the Sudakov form factor
makes on them is of higher order significance.  This argument holds a
fortiori given that the Sudakov form factor is introduced with compensating
terms, multiplying all of the Born cross section, cancelling its effects at
$\mathcal{O}\left(\as\right)$, effectively deferring them to yet lower
regions in the $\pTKH$ spectrum, i.e.~to regions where the $\pTKH^2/Q^{2}$
suppression of the regular cross section have already rendered its
contribution small. Finally, by way of a more quantitative ratification of
this claim we note the following integral,
\begin{equation}
\int_{\Lambda^{2}}^{Q^{2}}\frac{\mathd
  q^{2}}{q^{2}}\,\frac{q^{2}}{Q^{2}}\,\as \left(1+\as \,
A\,\log^{2}\frac{Q^{2}}{q^{2}}\right)\exp\left[-2\int_{q^{2}}^{Q^{2}}\frac{\mathd\mu^{2}}{\mu^{2}}\,
  A\,\as\,\log\frac{Q^{2}}{\mu^{2}}\right]=\as+\mathcal{O}\left(\as^{3}\right)\,,
\end{equation}
 performed assuming the coupling to be fixed at a scale $\mathcal{O}\left(Q\right)$.
This integral has the form of those which would be encountered in
evaluating the $\mathd \sigma_{{\scriptscriptstyle \mathcal{MF}}}$, showing
it to be equivalent to the analogous $\mathd \sigma_{{\scriptscriptstyle \mathcal{F}}}$
integral.\footnote{The same integral without the exponential or $\log^{2}Q^{2}/q^{2}$
terms.} 
Although we have resorted to a fixed coupling approximation here, experience
suggests this gives results compatible with other approaches, moreover, for
NLO accuracy we only require the $\mathd \sigma_{{\scriptscriptstyle
    \mathcal{F}}}$ and $\mathd \sigma_{{\scriptscriptstyle \mathcal{MF}}}$
integrations be the same up to $\mathcal{O}\left(\as^{2}\right)$ corrections
rather than the observed $\mathcal{O}\left(\as ^{3}\right)$.  We therefore
conclude that the contributions from $\mathd \sigma_{{\scriptscriptstyle
    \mathcal{F}}}$ and $\mathd \sigma_{{\scriptscriptstyle \mathcal{MF}}}$
may be safely taken as being equal.

With the correspondence between the $C_{ij}$ coefficient functions and the
NLO fixed order computations clear, as well as that between the
$\qT$ integral of $\mathd
\sigma_{{\scriptscriptstyle \mathcal{F}}}$ and $\mathd
\sigma_{{\scriptscriptstyle \mathcal{MF}}}$, we have thus proved that their
combination, viz.~the integrated / inclusive \MiNLO{} cross section,
is equivalent to that of a conventional fixed order computation at NLO
accuracy.

\subsection{Extension to vector-boson decay products}
\label{sub:extension-to-vector-boson-decays}

While our discussion has only referred explicitly to the rapidity
distributions of the $W$, $Z$ and Higgs bosons, the proof that these
distributions are NLO accurate straightforwardly extends to the case where
the cross section is further differential in the kinematics of their decay
products.  For the case of a scalar Higgs boson this is trivial, with the
decay essentially decoupled from the production stage, occurring
isotropically in its rest frame.

In the case of vector-boson production we are not aware of explicit results
in the literature for the $\pTKH \rightarrow 0$ limit of the NLO cross
sections, including the dependence on the momenta of the decay
products. However, we may infer such expressions using
ref.~\cite{Catani:2010pd}, which clarifies how to incorporate variables
parametrizing the momenta of such particles in the $b$-space resummation
formalism. This is achieved by modifying the resummation formula of
ref.~\cite{Collins:1984kg} via the insertion of the unit normalised LO
partonic cross section, differential in the decay variables, at the front of
the expression (outside the $b$-space integration), together with the
inclusion of a hard function, $H_{c}^{F}$, multiplying the coefficient
functions $C_{ij}$. The hard function partly encodes the NLO corrections to
the leading order distribution of the decay products' momenta. Clearly, with
these localised changes, if one carries out a perturbative expansion of this
$b$-space formula one will re-obtain the asymptotic $\pTKH \rightarrow 0$
cross sections of refs.~\cite{Arnold:1990yk,Glosser:2002gm}, modified by an
overall factor of the leading order cross section differential in the decay
variables and with the $\mathcal{O}\left(\as\right)$ terms in the $C_{ij}$
coefficient functions modified to include NLO corrections to that
distribution. Thus including dependencies on the details of the vector-boson
decay leaves our proof unmodified up to inclusion of an overall angular
factor describing the decay kinematics and, in general, an implicit
dependence on these variables in the $\mathcal{O}\left(\as\right)$ part of
the coefficient functions.

Taking a different tack, for the case that the decay variables are the
Collins-Soper angles~\cite{Collins:1977iv}, this extension of the resummation
formalism was already well studied in transverse-momentum
space~\cite{Ellis:1997ii} and impact-parameter space~\cite{Balazs:1997xd,
  Ellis:1997sc}.  In the former article, in transverse momentum space, one
can readily see the angular dependence of the resummed cross section present
as a global angular prefactor. In fact, in refs.~\cite{Ellis:1997sc,
  Balazs:1997xd, Ellis:1997ii} the only modification of the `standard'
resummation formulae is just that: the coefficient functions are not modified
with respect to those used for resummation of vector-boson production
processes where the decay is not considered. We attribute this to the fact
that the virtual QCD corrections, simply act to rescale the hadronic tensor
for the production vertex by an overall factor, i.e.~the dependence on the
Collins-Soper decay angles (defined in the decaying boson's rest frame) is
the same in the virtual and Born contributions to the cross
section. Moreover, only those parts of the real cross section which are
singular when $\pTKH \rightarrow0$ are integrated over in combining with the
virtual corrections to give the NLO $C_{ij}$ terms, as one would expect,
these parts therefore have an angular dependence for the decay products which
is the same as that of the Born term.%
\footnote{This is elaborated on in Appendices~B and~E of ref.~\cite{Balazs:1997xd}.}
Consequently, the only modification needed to the asymptotic $\pTKH $
spectrum formula in ref.~\cite{Arnold:1990yk}, to account for these
angular variables, is the inclusion of an overall factor describing
the dependence at the relative $\mathcal{O}\left(\as^{0}\right)$ / Born
level.

\section{Implementation and plots}
\label{sec:pheno}
In the following we discuss results obtained from an implementation of our
revised \MINLO{} method in the \POWHEGBOX{}, for the
\HJ{}~\cite{Alioli:2008tz, Campbell:2012am}, \WJ{} and \ZJ{}
generators~\cite{Alioli:2008gx, Alioli:2010qp}. For the purpose of this
study, we will use throughout the pdf set MSTW2008NLO~\cite{Martin:2009iq}.
Any other popular set~\cite{Lai:2010vv, Ball:2012cx} can be used
equivalently, and we are not interested here to explore pdf dependencies.
For the Higgs boson case, we use throughout $\MH=125\,$GeV and in the \H{}
generator we set the {\tt hfact} parameter to 100~\cite{Dittmaier:2012vm}.

We have advocated that the \MINLO{} prescription, updated as described in the
previous section, can achieve NLO$^{(0)}$ accuracy, that is to say, it can
describe inclusive boson distributions at NLO accuracy. We thus begin with
the most inclusive quantity, i.e.~the total cross section. We would like to
see if the total cross section obtained by integrating the \BJ-\MiNLO{}
formulae is compatible with the NLO total cross section obtained with
standard NLO calculations. Notice that we expect agreement only up to terms
of higher order in $\as$, since the \BJ-\MiNLO{} results include terms of
higher order, and also since the meaning of the scale choice is different in
the two approaches. For similar reasons, we do not expect the scale variation
bands to be exactly the same in the two approaches.

In tab.~\ref{tab:totH}, we list the results for the total cross sections
obtained with the \HJ{}-\MiNLO{} and the \H{} programs, both at full NLO
level and at leading order, for different scale combinations.  The scale
variation in the \HJ{}-\MiNLO{} result is obtained by multiplying the
factorization scale and each of the several renormalization scales that
appear in the procedure by the scale factors $\KFA$ and $\KRA$,
respectively. The Sudakov form factor is also changed according to the
prescription of Appendix~\ref{app:scalevar}.  The maximum and minimum of the
cross-sections are highlighted.  In the case of \MiNLO{}, the leading-order
result is obtained by keeping only the Born term in \POWHEG{} (i.e.~by
setting the {\tt bornonly} flag to 1), and by downgrading the Sudakov form
factor to pure NLL accuracy, i.e.~we set $B_2$ to zero.

In the \MiNLO{} case, the central value is chosen according to the procedure
discussed earlier, with more than one renormalization scale for each phase
space point. In the \H{} fixed order calculation, we choose as central
renormalization and factorization scales the boson mass.
\begin{table}[htb]
\centering
{\small
\begin{tabular}{|c|c|c|c|c|c|c|c|c|c|}
\hline
\multicolumn{8}{|c|}{Higgs boson production total cross sections in pb at the LHC, 8~TeV}\\ \hline
 $K_R,K_F$  & $1,1$ & $1,2$ & $2,1$ & $1,\half$ & $\half,1$ & $\half,\half$ & $2,2$ \\ \hline
\HJ{}-\MiNLO{} NLO   & $ 13.33(3)$ & $ 13.49(3)$ & $ \mathbf{11.70(2)}$ & $ 13.03(3)$ & $ \mathbf{16.53(7)}$ & $ 16.45(8)$ & $ 11.86(2)$\\ \hline
\H{} NLO   & $ 13.23(1)$ & $ 13.28(1)$ & $ \mathbf{11.17(1)}$ & $ 13.14(1)$ & $ \mathbf{15.91(2)}$ & $ 15.83(2)$ & $ 11.22(1)$\\ \hline
\HJ{}-\MiNLO{} LO &$ 8.282(7)$ & $ 8.400(7)$ & $ \mathbf{5.880(5)}$ & $ 7.864(6)$ & $ \mathbf{18.28(2)}$ & $ 17.11(2)$ & $ 5.982(5)$
 \\ \hline
\H{} LO & $ 5.741(5)$ & $ 5.758(5)$ & $ \mathbf{4.734(4)}$ & $ 5.644(5)$ & $ \mathbf{7.117(6)}$ & $ 6.996(6)$ & $ 4.748(4)$ \\ \hline
 \end{tabular}
}
\caption{Total cross section for Higgs boson production at the 8~TeV LHC,
  obtained with the \HJ{}-\MiNLO{} and the \H{} programs, both at full NLO
  level and at leading order, for different scales combinations. The maximum
  and minimum are highlighted.}
\label{tab:totH} 
\end{table}
From the table, it is clear that the standard NLO result and the integrated
\HJ{}-\MiNLO{} one are fairly consistent, both at the NLO and at the LO
level.  At the NLO level, the renormalization-scale variation dominates the
uncertainty band, and it turns out to be very similar for the \HJ{}-\MiNLO{}
and \H{} results, with the first one being slightly shifted upwards. The
central values are even closer.  Notice that the factorization scale
variation is wider for the \HJ{}-\MiNLO{} result, a fact that we will comment
on later.

At leading order the \HJ-\MiNLO{} central result exceeds the fixed order one
by almost 50\%.  We again see that the renormalization scale variation
dominates the uncertainties.  The scale variation, however, is quite larger
than that of the fixed order result.

For $W^-$ production we have considered both the LHC at 8~TeV configuration
(tab.~\ref{tab:totW}) and the Tevatron at 1.96~TeV (tab.~\ref{tab:totW_TEV}).
\begin{table}[htb]
\centering
{\small
\begin{tabular}{|c|c|c|c|c|c|c|c|c|c|}
\hline
\multicolumn{8}{|c|}{$W^-\to e^- \bar{\nu}$ production total cross sections in nb at the LHC, 8~TeV}\\ \hline
 $K_R,K_F$  & $1,1$ & $1,2$ & $2,1$ & $1,\half$ & $\half,1$ & $\half,\half$ & $2,2$ \\ \hline
\WJ{}-\MiNLO{} NLO &      $4.35(1)$&$4.65(1)$&$4.031(7)$&$\mathbf{3.818(8)}$&$\mathbf{4.84(2)}$&$4.62(2)$&$4.462(8)$ \\ \hline
\W{} NLO & $ 4.612(8)$ & $ \mathbf{4.738(8)}$ & $ 4.552(8)$ & $ \mathbf{4.425(7)}$ & $ 4.687(8)$ & $ 4.530(8)$ & $ 4.703(8)$ \\ \hline
\WJ{}-\MiNLO{} LO & $3.182(1)$&$3.862(1)$&$2.713(1)$&$\mathbf{2.4531(1)}$&$\mathbf{5.006(2)}$&$3.792(2)$&$3.305(1)$
\\ \hline
\W{} LO & $ 4.002(6)$ & $ 4.379(7)$ & $ 3.999(6)$ & $ 3.566(6)$ & $ 3.999(6)$ & $ \mathbf{3.566(6)}$ & $ \mathbf{4.379(7)}$ \\ \hline
 \end{tabular}
}
\caption{Total cross section for $W^-\to e^- \bar{\nu}$ production at the
  8~TeV LHC, obtained with the \WJ{}-\MiNLO{} and the \W{} programs, both at
  full NLO level and at leading order, for different scales combinations. The
  maximum and minimum are highlighted.}
\label{tab:totW} 
\end{table}
\begin{table}[htb]
\centering
{\footnotesize
\begin{tabular}{|c|c|c|c|c|c|c|c|c|c|}
\hline
\multicolumn{8}{|c|}{$W^-\to e^- \bar{\nu}$ production total cross sections in nb at the Tevatron}\\ \hline
 $K_R,K_F$  & $1,1$ & $1,2$ & $2,1$ & $1,\half$ & $\half,1$ & $\half,\half$ & $2,2$ \\ \hline
\WJ{}-\MiNLO{} NLO & $1.204(2)$&$1.225(2)$&$\mathbf{1.155(1)}$&$1.177(2)$&$1.240(4)$&$\mathbf{1.288(4)}$&$1.186(1)$
 \\ \hline
\W{} NLO &  $ 1.253(1)$ & $ 1.264(1)$ & $ \mathbf{1.232(1)}$ & $ 1.246(1)$ & $ \mathbf{1.281(1)}$ & $ 1.273(1)$ & $ 1.241(1)$ \\ \hline
\WJ{}-\MiNLO{} LO &$0.9633(3)$&$0.9988(3)$&$\mathbf{0.8334(3)}$&$0.8962(3)$&$\mathbf{1.5971(6)}$&$1.4696(5)$&$0.8677(3)$
 \\ \hline
\W{} LO &$ 1.127(1)$ & $ \mathbf{1.130(1)}$ & $ 1.127(1)$ & $ \mathbf{1.116(1)}$ & $ 1.127(1)$ & $ 1.116(1)$ & $ 1.130(1) $ \\ \hline
 \end{tabular}
}
\caption{Total cross section for $W^-\to e^- \bar{\nu}$ production at the
  1.96~TeV Tevatron, obtained with the \WJ{}-\MiNLO{} and the \W{} programs,
  both at full NLO level and at leading order, for different scales
  combinations. The maximum and minimum are highlighted.}
\label{tab:totW_TEV}
\end{table}
Here we notice that the \WJ-\MiNLO{} NLO result has a much wider
scale-variation band than the fixed-order one. In both cases, the band is
larger by about a factor of 3.  The central value is lower in both cases by
about 4-5\%. In the leading order case, the \WJ-\MiNLO{} scale band is more
than twice as large as the fixed order one at the LHC.  At the Tevatron, the
scale variation for the \W{} LO result is clearly too small, the NLO result
being incompatible with it.  On the other hand, for both LHC and Tevatron
predictions, if only symmetric scale variations are considered (i.e.~the last
two columns of the tables), the \WJ-\MiNLO{} NLO band contains the fixed
order one, and is comparable to it.

We notice now, that comparing full independent scale variation in the
\WJ-\MiNLO{} and in the \W{} approaches does not seem to be totally fair. In
fact, in the \W{} case, there is no renormalization scale dependence at LO,
while there is such a dependence in \WJ-\MiNLO{}. Since the \W{} and
\WJ-\MiNLO{} cross sections are related in a known way, we can track the
effects of scale variation in both formulae, at least in the LO case. We
write, schematically, the leading order $W$ production cross section as
\begin{equation}
\label{eq:sigchem}
\sigma\(\KFA\)=\left[ {\cal L}\(\KFA^2 Q^2\) \otimes \sigma_0\right] ,
\end{equation}
where with the convolution sign we schematically represent the integral with
the luminosity ${\cal L}$.  We have explicitly indicated the factorization
scale dependence in the formula. Here $Q$ stands for the reference scale,
that, in this case, is the $W$ mass. In order to make contact with the LO
\MiNLO{} formula, we rewrite eq.~(\ref{eq:sigchem}) as
\begin{equation}
\label{eq:sigchem1}
\sigma\(\KFA\)=\int_0^{Q^2} \mathd q^2 \frac{\mathd}{\mathd q^2}   \exp{\cal S}(Q,q)
\left[{\cal L}\(\KFA^2 q^2\)\otimes\sigma_0 \right]\,,
\end{equation}
where
\begin{equation}
{\cal S}(Q,q)=-\int_{q^2}^{Q^2} \frac{\mathd \mu^2}{\mu^2} \as\(\mu^2\)
\left(A_1 \log\frac{Q^2}{\mu^2}+B_1\right)\,,
\end{equation}
that, being the integral of an exact differential, yields
eq.~(\ref{eq:sigchem}) up to non-perturbative effects, that we here neglect,
related to the low-end of the integration. Taking explicitly the derivative,
we get
\begin{eqnarray}
\sigma\(\KFA\)&=&
\int_0^{Q^2} \frac{\mathd q^2}{q^2}  \exp{\cal S}(Q,q)
  \bigg\{ \as\(\KFA^2 q^2\) \left[P\otimes {\cal L}\(\KFA^2 q^2\)\right]
\nonumber \\
&& {}
+\as\(q^2\)\( A_1\log\frac{Q^2}{q^2}+B_1 \) {\cal L}\(\KFA^2 q^2\) \bigg\}\otimes \sigma_0
\nonumber \\
&=& \int_0^{Q^2} \frac{\mathd q^2}{q^2}  \exp{\cal S}(Q,q)
\bigg\{\as\(\KFA^2 q^2\) \left[\hat{P}\otimes {\cal L}\(\KFA^2 q^2\)\right]
\otimes \sigma_0 
\nonumber \\
&& {} + \left[ \as\(q^2\)-\as\(\KFA^2 q^2\) \right] \left(A_1
\log\frac{Q^2}{q^2}+B_1\right) {\cal L}\(\KFA^2 q^2\)\otimes 
\sigma_0 \bigg\}\;.
\end{eqnarray}
We have made use of the schematic equation
\begin{equation}
\left[P\otimes {\cal L}\(\KFA^2 q^2\)\right]+\left[A_1\log\frac{Q^2}{q^2}+B_1\right] {\cal L}\(\KFA^2 q^2\)
=\left[\hat{P}\otimes {\cal L}\(\KFA^2 q^2\)\right]\;, 
\end{equation}
where $P$ stand for the regularized (i.e.~including the plus prescriptions)
Altarelli-Parisi splitting functions, while $\hat{P}$ are the unregularized
ones. In the last case, we assume that the integration in the convolution has
the correct kinematic cutoff.

On the other hand, the \MiNLO{} formula for the total cross section has the form
\begin{equation}
\sigma\(\KRA,\KFA\)= \int_0^{Q^2} \frac{\mathd q^2}{q^2} \as\(\KRA^2 q^2\) \left\{
\left[\hat{P}\otimes {\cal L}\(\KFA^2 q^2\)\right] \exp{\cal S}(Q,q) 
\otimes \sigma_0 +  {\cal L}\(\KFA^2 q^2\) \otimes R_f \right\}\,,
\end{equation}
where the last term corresponds to the $R_f$ finite contribution. We thus see
that an independent scale variation in the \W{} formula corresponds at least
in part to a symmetric scale variation in the \MiNLO{} formula. It is thus
not surprising that the \MiNLO{} independent scale variation is so much
larger than the \W{} one also at NLO. If we limit ourselves to consider only
symmetric scale variations, the \MiNLO{} and the \W{} results are more
consistent, although the \W{} scale variation band is extremely small. The
full NNLO result for the $W$ cross section, on the other hand, lies outside
the \W{} uncertainty band in this case, which suggests that, after all, the
\MiNLO{} scale band is not unreasonable.

In tables~\ref{tab:totZ_TEV}, \ref{tab:totZ} and~\ref{tab:totZ_LHC14} we show
the total cross section for $Z\to e^+e^-$ production at the Tevatron and at
the LHC at 8 and 14~TeV.
\begin{table}[htb]
\centering
{\scriptsize
\begin{tabular}{|c|c|c|c|c|c|c|c|c|c|}
\hline
\multicolumn{8}{|c|}{$Z\to e^+ e^-$ production total cross sections in nb at the Tevatron}\\ \hline
 $K_R,K_F$  & $1,1$ & $1,2$ & $2,1$ & $1,\half$ & $\half,1$ & $\half,\half$ & $2,2$ \\ \hline
\ZJ{}-\MiNLO{} NLO & $0.2407(4)$&$0.2434(3)$&$\mathbf{0.2321(3)}$&$0.2378(3)$&$0.2459(8)$&$\mathbf{0.2552(7)}$&$0.2359(3)$\\ \hline
\Z{} NLO &$ 0.2498(3)$ & $ 0.2513(3)$ & $ \mathbf{0.2455(3)}$ & $ 0.2491(3)$ & $ \mathbf{0.2552(3)}$ & $ 0.2543(3)$ & $ 0.2467(3)$
 \\ \hline
\ZJ{}-\MiNLO{} LO &$0.19508(7)$&$0.19883(7)$&$\mathbf{0.16948(6)}$&$0.18572(6)$&$\mathbf{0.3213(1)}$&$0.3029(1)$&$0.17336(6)$ \\ \hline 
\Z{} LO & $ \mathbf{0.2052(2)}$ & $ \mathbf{0.2041(2)}$ & $ 0.2052(2)$ & $ 0.2052(2)$ & $ 0.2052(2)$ & $ 0.2052(2)$ & $ 0.2041(2)$ \\ \hline
 \end{tabular}
}
\caption{ Total cross section for $Z^-\to e^+ e^-$ production, obtained with
  the \ZJ{}-\MiNLO{} and the \Z{} programs, both at full NLO level and at
  leading order, for different scales combinations. The maximum and minimum
  are highlighted.}
\label{tab:totZ_TEV}
\end{table}
\begin{table}[htb]
\centering
{\footnotesize
\begin{tabular}{|c|c|c|c|c|c|c|c|c|c|}
\hline
\multicolumn{8}{|c|}{$Z\to e^+ e^-$ production total cross sections in nb at the LHC, 8~TeV}\\ \hline
 $K_R,K_F$  & $1,1$ & $1,2$ & $2,1$ & $1,\half$ & $\half,1$ & $\half,\half$ & $2,2$ \\ \hline
\ZJ{}-\MiNLO{} NLO & $1.044(2)$&$1.111(2)$&$0.974(1)$&$\mathbf{0.933(1)}$&$\mathbf{1.157(3)}$&$1.116(3)$&$1.067(1)$
 \\ \hline
\Z{} NLO & $ 1.107(1)$ & $ \mathbf{1.135(1)}$ & $ 1.093(1)$ & $ \mathbf{1.066(1)}$ & $ 1.125(1)$ & $ 1.091(1)$ & $ 1.126(1)$ \\ \hline
\ZJ{}-\MiNLO{} LO & $0.7751(3)$&$0.9260(3)$&$0.6632(2)$&$\mathbf{0.6111(2)}$&$\mathbf{1.2114(4)}$&$0.9405(3)$&$0.7950(3)$
\\ \hline
\Z{} LO & $ 0.9597(9)$ & $ \mathbf{1.0443(1)}$ & $ 0.9604(9)$ & $ \mathbf{0.8647(9)}$ & $ 0.9604(9)$ & $ 0.8647(9)$ & $ 1.0443(1)$ \\ \hline
 \end{tabular}
}
\caption{Total cross section for $Z^-\to e^+ e^-$ production, obtained with
  the \ZJ{}-\MiNLO{} and the \Z{} programs, both at full NLO level and at
  leading order, for different scales combinations. The maximum and minimum
  are highlighted.}
\label{tab:totZ} 
\end{table}
\begin{table}[htb]
\centering
{\scriptsize
\begin{tabular}{|c|c|c|c|c|c|c|c|c|c|}
\hline
\multicolumn{8}{|c|}{$Z\to e^+ e^-$ production total cross sections in nb at the 14~TeV LHC}\\ \hline
 $K_R,K_F$  & $1,1$ & $1,2$ & $2,1$ & $1,\half$ & $\half,1$ & $\half,\half$ & $2,2$ \\ \hline
\ZJ{}-\MiNLO{} NLO &$1.916(5)$&$2.065(6)$&$1.776(2)$&$\mathbf{1.662(3)}$&$\mathbf{2.18(1)}$&$2.022(6)$&$1.987(3)$
\\ \hline
\Z{} NLO &$ 2.039(3)$ & $ \mathbf{2.100(3)}$ & $ 2.015(2)$ & $ \mathbf{1.938(2)}$ & $ 2.068(3)$ & $ 1.984(2)$ & $ 2.092(3)$ \\ \hline
\ZJ{}-\MiNLO{} LO &$1.3827(5)$&$1.7322(6)$&$1.1806(4)$&$\mathbf{1.0348(3)}$&$\mathbf{2.1280(7)}$&$1.5677(5)$&$1.4831(5)$
 \\ \hline
\Z{} LO &$ 1.793(2)$ & $ \mathbf{2.014(2)}$ & $ 1.793(2)$ & $ 1.555(2)$ & $ 1.793(2)$ & $ \mathbf{1.555(2)}$ & $ 2.014(2) $ \\ \hline
 \end{tabular}
}
\caption{ Total cross section for $Z^-\to e^+ e^-$ production, obtained with
  the \ZJ{}-\MiNLO{} and the \Z{} programs, both at full NLO level and at
  leading order, for different scales combinations. The maximum and minimum
  are highlighted.}
\label{tab:totZ_LHC14}
\end{table}
As in the $W$ case, we see that the full 7-point scale variation yields a
much wider band in the \ZJ-\MiNLO{} result with respect to the \Z{} one.

\begin{figure}[htb]
\begin{center}
\includegraphics[width=0.49\textwidth]{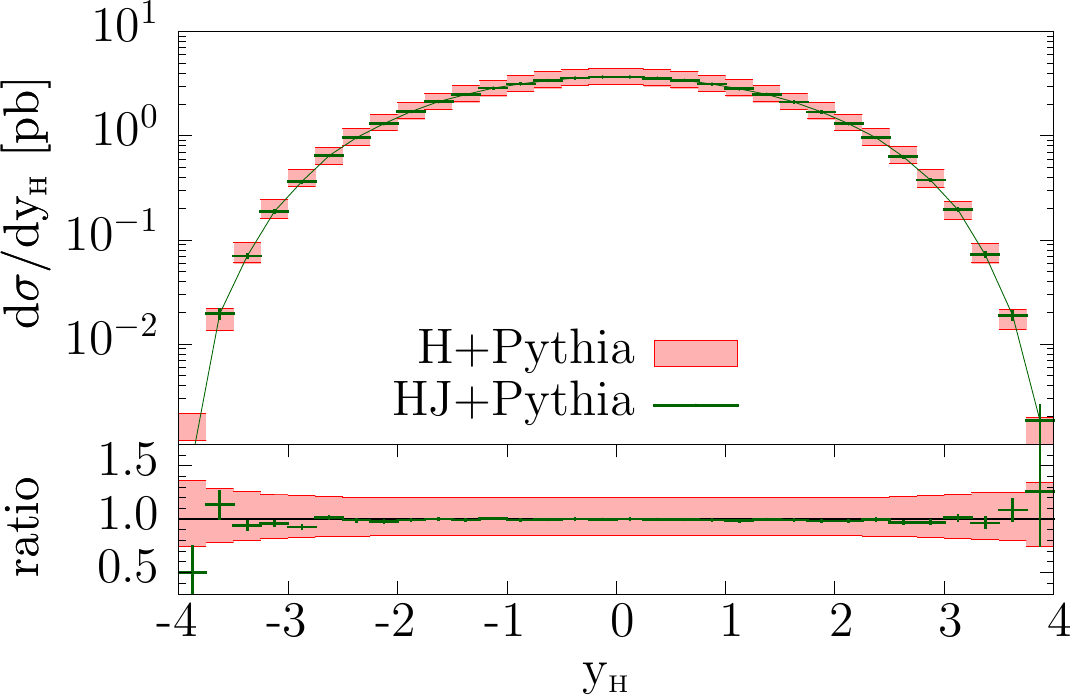}
\includegraphics[width=0.49\textwidth]{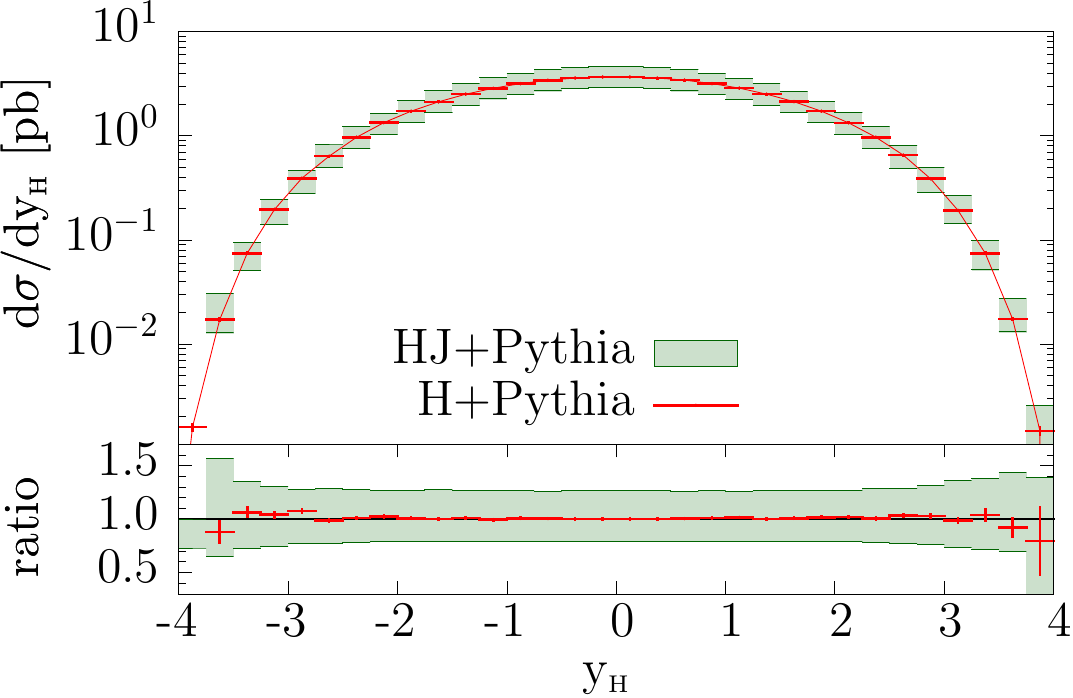}
\end{center}
\caption{Comparison between the \H+\PYTHIA{} result and the
  \HJ-\MiNLO+\PYTHIA{} result for the Higgs-boson rapidity distribution at
  the LHC at 8~TeV.  The left plot shows the 7-point scale-variation band for
  the \H{} generator, while the right plot shows the \HJ-\MiNLO{} 7-point
  band.}
\label{fig:Hy} 
\end{figure}
After having shown that the total cross sections obtained with the
\BJ{}-\MiNLO{} generator are in good agreement with the standard NLO cross
sections, we would like to show that also the rapidity distributions are in
good agreement. We thus show in fig.~\ref{fig:Hy} the rapidity distribution
of the Higgs boson at the 8~TeV LHC, computed with the \H{} and with the
\HJ{}-\MiNLO{} generators, both interfaced to
\PYTHIA~6~\cite{Sjostrand:2006za} for shower. We have used the Perugia-0 tune
of \PYTHIA{} (that is to say, {\tt PYTUNE(320)}). Hadronization, underlying
event and multi-parton collisions were turned off.  The two plots show the
scale-variation band for each generator. The band is obtained as the upper
and lower envelope of the results obtained by setting the scale factor
parameters $(\KRA,\KFA)$ to $(1,1)$, $(1,2)$, $(2,1)$, $(1,\half)$,
$(\half,1)$, $(\half,\half)$ and $(2,2)$.  We see considerable agreement
between the two approaches, with the scale-variation band of the \HJ-\MiNLO{}
result being slightly larger.

\begin{figure}[htb]
\begin{center}
\includegraphics[width=0.49\textwidth]{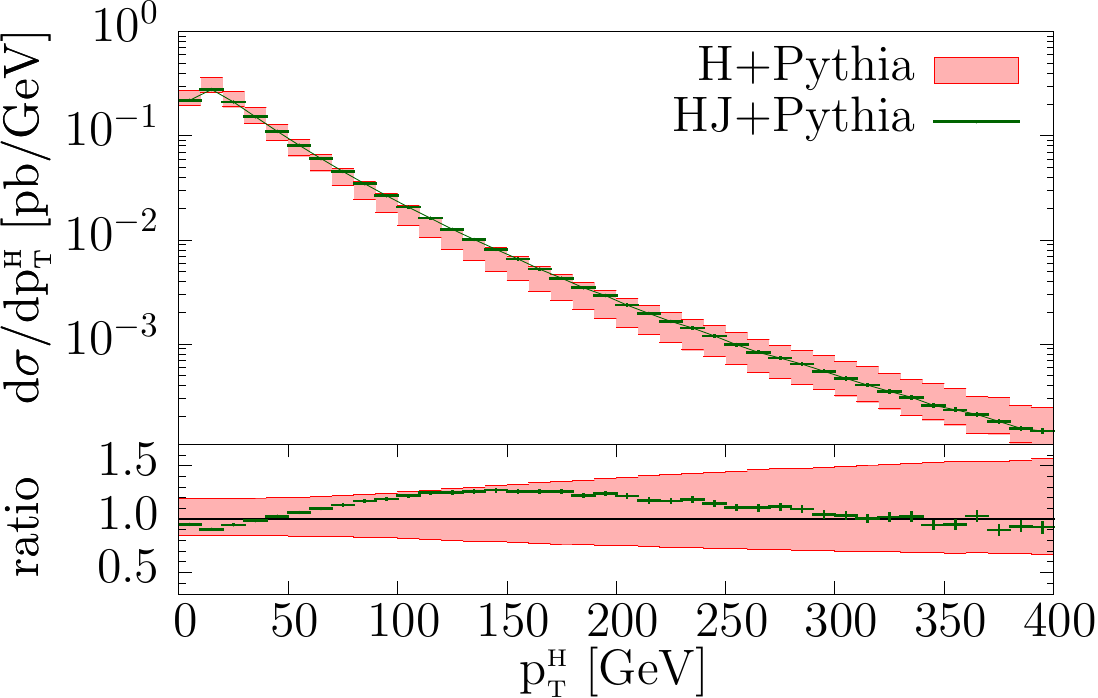}
\includegraphics[width=0.49\textwidth]{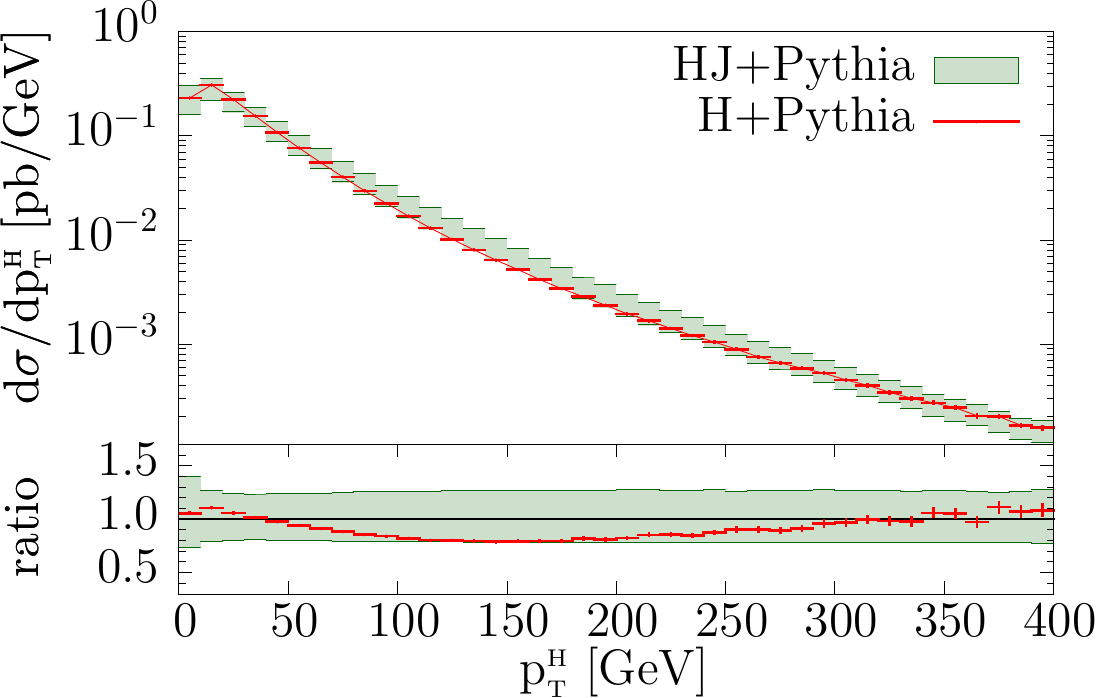}
\end{center}
\caption{Comparison between the \H+\PYTHIA{} result and the
  \HJ{}-\MiNLO{}+\PYTHIA{} result for the Higgs boson transverse-momentum
  distribution. The bands are obtained as in fig.~\ref{fig:Hy}.}
\label{fig:Hpt} 
\end{figure}
\begin{figure}[htb]
\begin{center}
\includegraphics[width=0.49\textwidth]{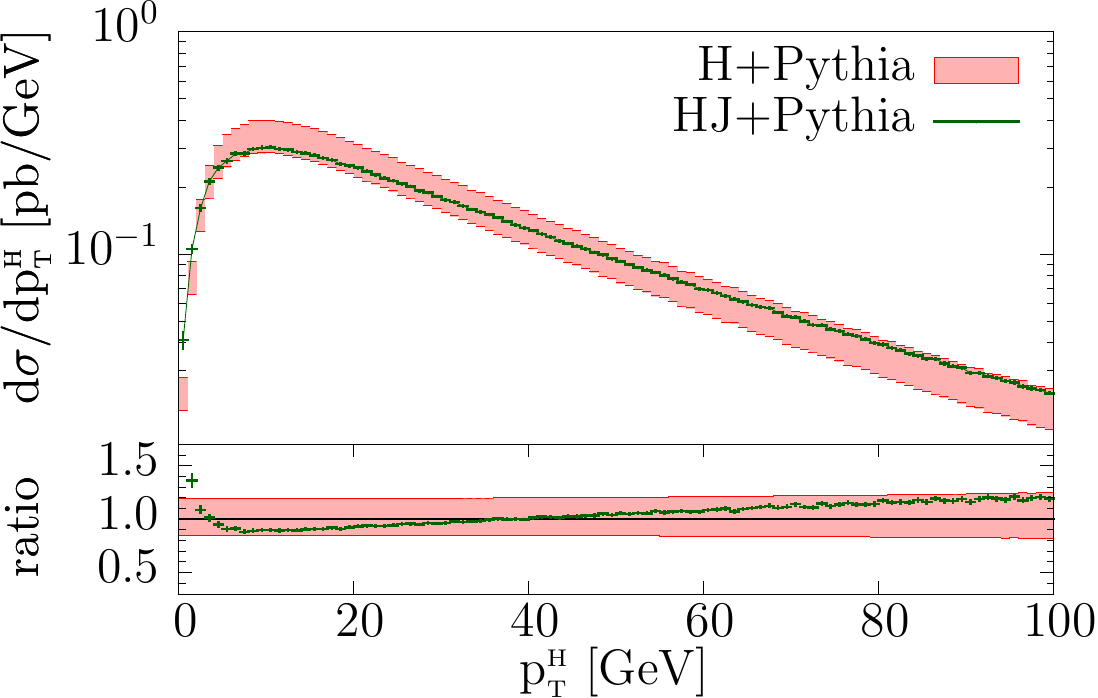}
\includegraphics[width=0.49\textwidth]{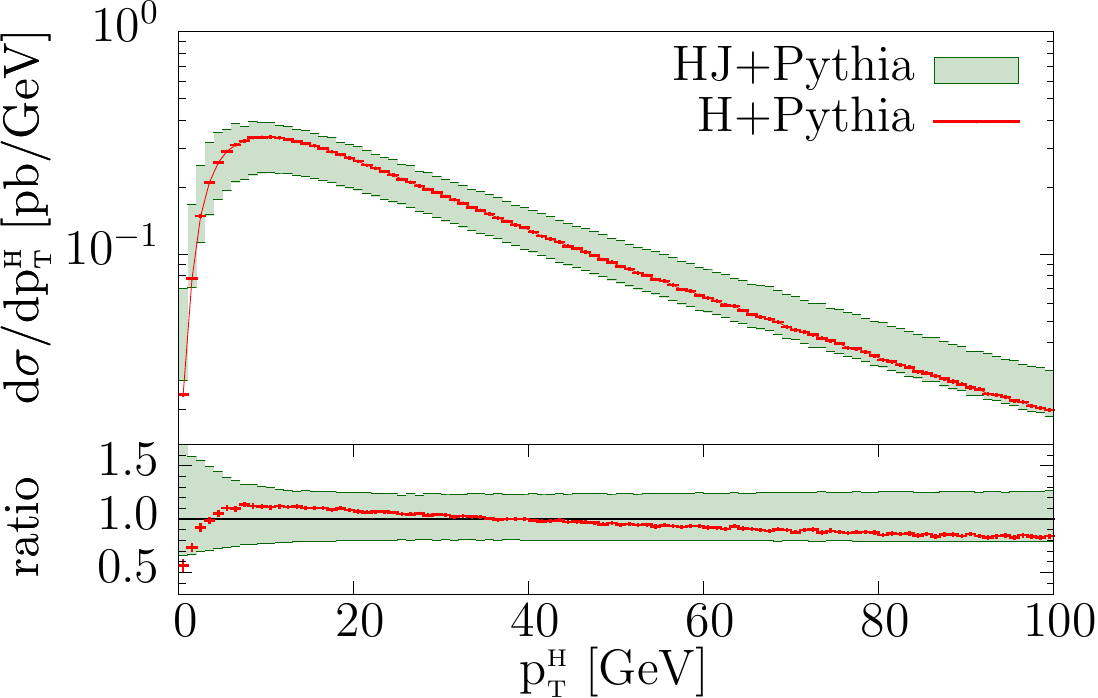}
\end{center}
\caption{Same as fig.~\ref{fig:Hpt} for a different $\pt^{\rm\sss H}$ range.}
\label{fig:Hptzm} 
\end{figure}
In figs.~\ref{fig:Hpt} and~\ref{fig:Hptzm} we show the Higgs transverse
momentum distributions.  We begin by noticing that the central values of the
\H{} and \HJ-\MiNLO{} generators are in very good agreement.  This is not a
surprise, since in the \H{} generator, the parameter {\tt hfact}, that
separates the real cross section contribution into the sum of a singular and
a finite one, was set to the value $\MH/1.2$, motivated by the fact that this
yields better agreement with the NNLO result.

We notice that, for large transverse momenta, the \HJ-\MiNLO{} generator has
a smaller scale variation band with respect to the \H{} one. We expect this
behaviour, since the \HJ-\MiNLO{} generator achieves NLO accuracy for one-jet
inclusive distributions, while the \H{} generator is only tree-level
accurate.  We also notice that the scale uncertainty band of \HJ-\MiNLO{}
widens at small transverse momentum. This behaviour is also expected, since,
in that direction, we approach the strong coupling regime. Observe also that
the \H{} result does not show a realistic scale uncertainty in the
$\pt^{\sss\rm H}<\MH$ region.  This too is understood, and it follows from
the fact that this region is dominated by $S$-type events (see
refs.~\cite{Dittmaier:2012vm, Nason:2012pr} for a detailed explanation).

As a last point, we see from fig.~\ref{fig:Hptzm}, that a noticeable
difference in shape is present in the very small transverse-momentum
region. This again does not come as a surprise, since the \POWHEG-generated
Sudakov form factor in the \H{} generator differs by NNLL terms, and also by
non-singular contributions, from the \HJ-\MiNLO{} one. Notice also that,
unlike in the \H{} case~\cite{Nason:2012pr}, the scale variation in the
\HJ-\MiNLO{} generator induces a change in shape of the transverse momentum
spectrum in the Sudakov region, leading to a better understanding of the
associated uncertainty.

\begin{figure}[htb]
\begin{center}
\includegraphics[width=0.49\textwidth]{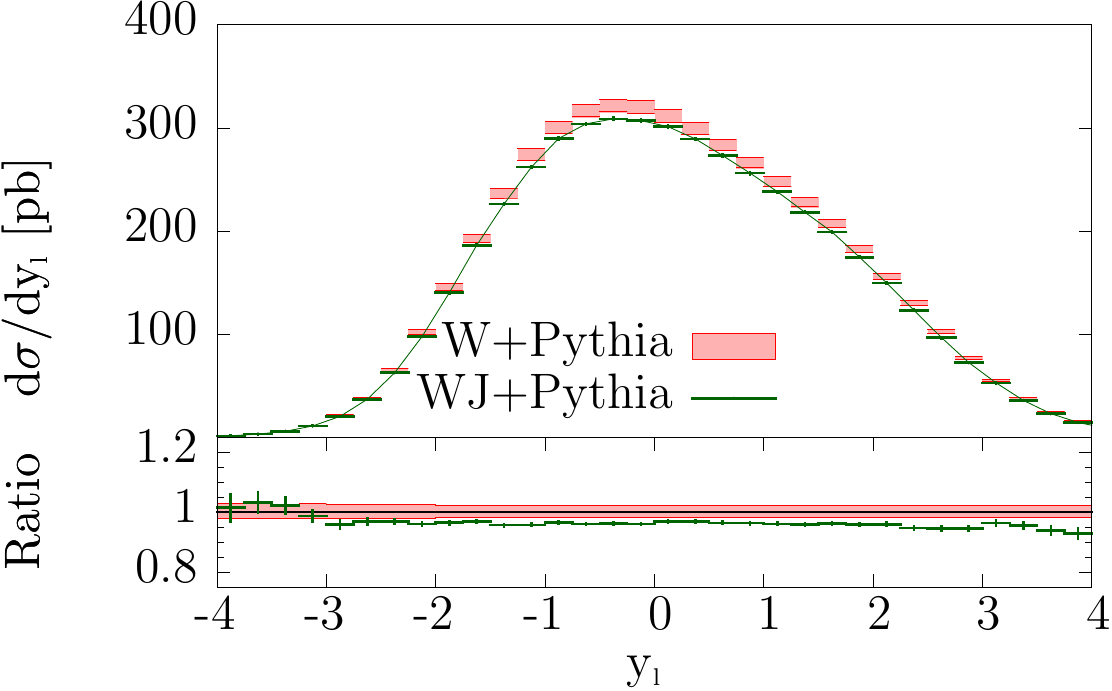}
\includegraphics[width=0.49\textwidth]{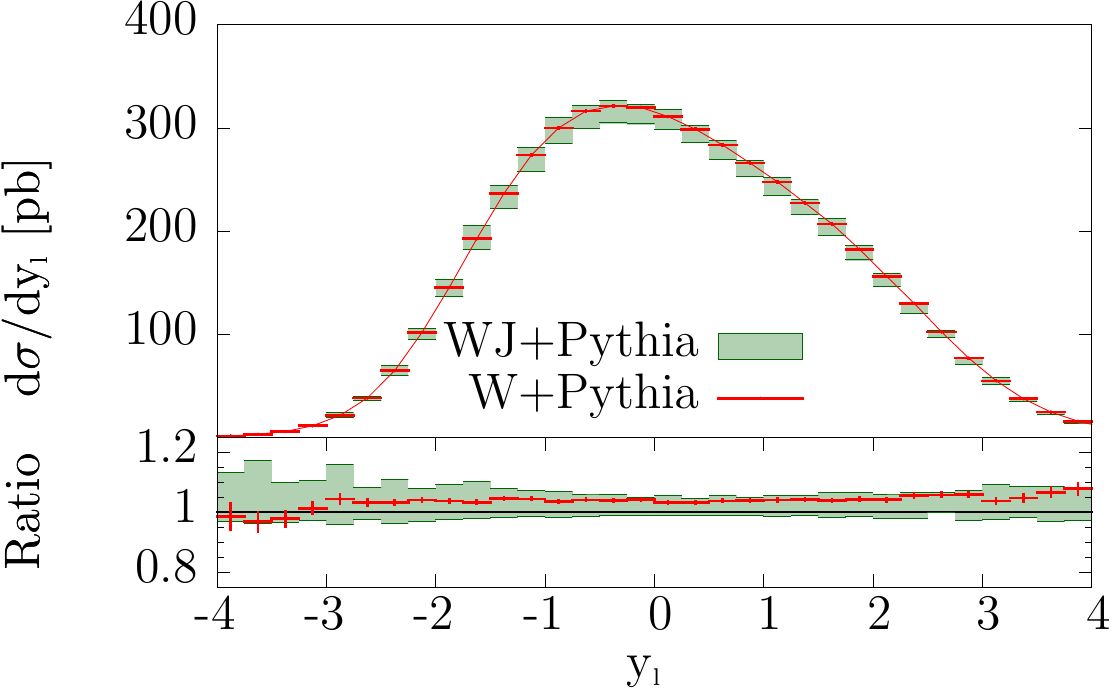}
\end{center}
\caption{Comparison between the \W+\PYTHIA result and the
  \WJ-\MiNLO+\PYTHIA{} result for the $l^-$ rapidity distribution at the
  Tevatron. The bands are obtained as in fig.~\ref{fig:Hy} for the
  \W{}+\PYTHIA{} generator, while for the \WJ-\MiNLO+\PYTHIA{} generator they
  are obtained by taking the upper and lower envelope of the curves computed
  with $\KRA=\KFA=\{1/2,1,2\}$.}
\label{fig:Wy} 
\end{figure}
We now turn to the case of $W^-$ production. Motivated by the discussion
given for the total cross section case, we consider only a 3-point scale
variation, i.e.~$\KRA=\KFA=\{1/2,1,2\}$ for the \WJ-\MiNLO{} generator.  In
fig.~\ref{fig:Wy} we show the $l^-$ rapidity distribution at the Tevatron
computed with the \W{} and \WJ{}+\MINLO{} generators. We essentially see no
shape difference in this distribution, therefore, as for the inclusive cross
section, we find that the \WJ{}+\MINLO{} central value is about 5\% below the
\W{} one. The \WJ{} band is slightly larger than the \W{} one for central
rapidities, widening towards larger rapidities.

\begin{figure}[htb]
\begin{center}
\includegraphics[width=0.49\textwidth]{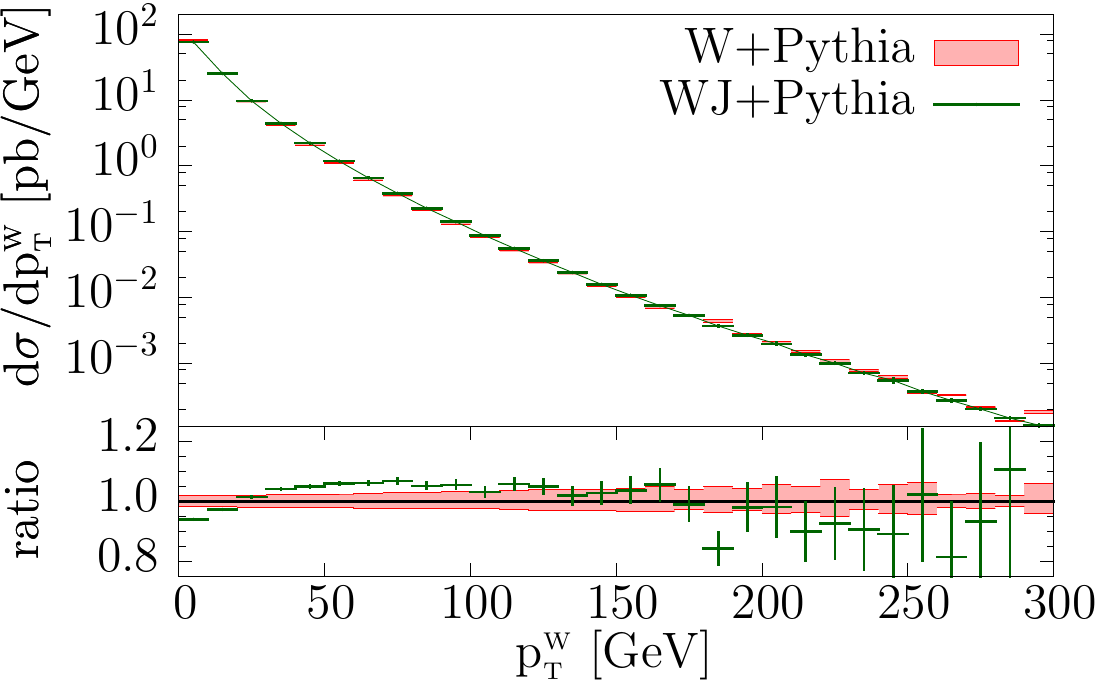}
\includegraphics[width=0.49\textwidth]{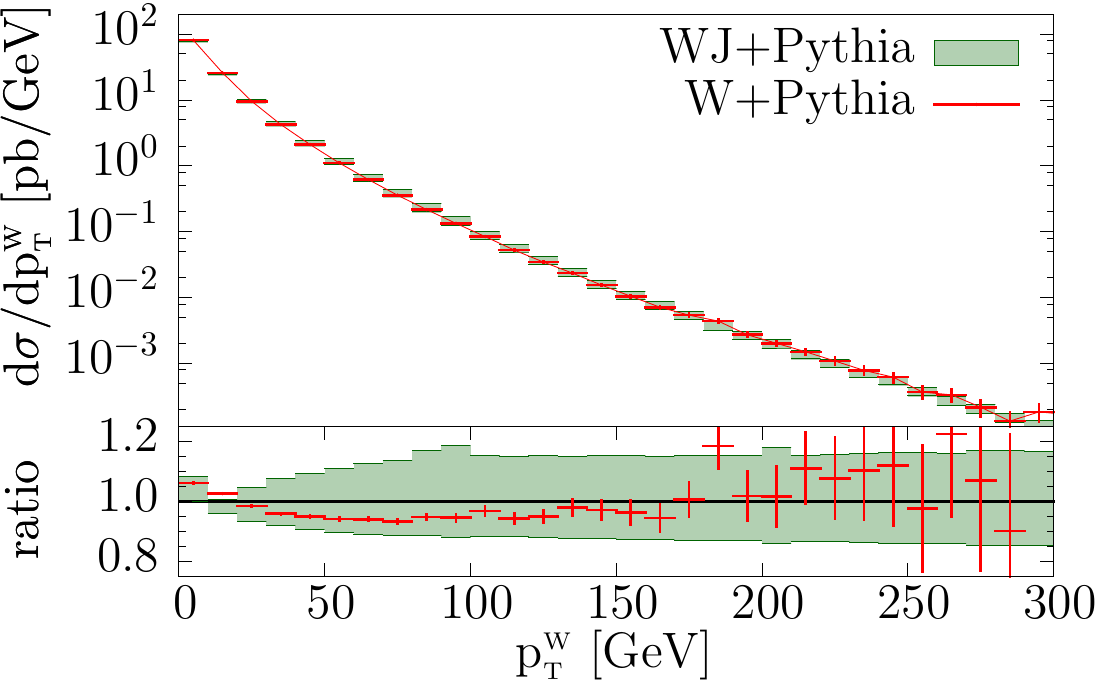}
\end{center}
\caption{Comparison between the \W+\PYTHIA result and the
  \WJ-\MiNLO+\PYTHIA{} result for the $W^-$ transverse-momentum distribution
  at the Tevatron. The bands are obtained as in fig.~\ref{fig:Wy}.}
\label{fig:Wpt} 
\end{figure}
\begin{figure}[htb]
\begin{center}
\includegraphics[width=0.49\textwidth]{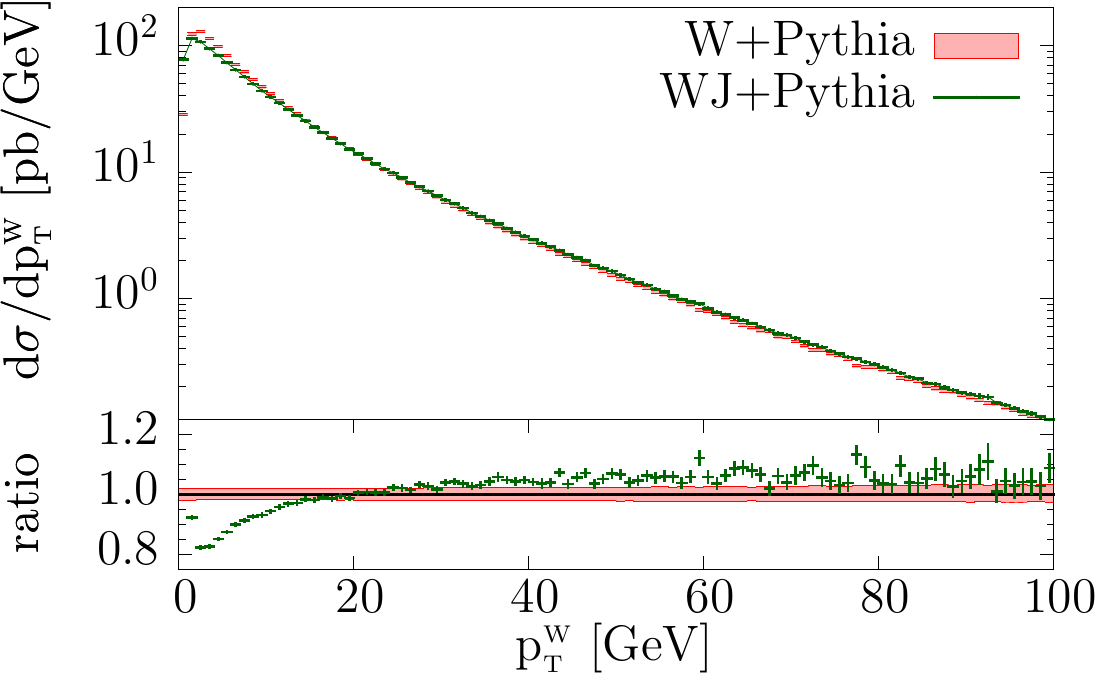}
\includegraphics[width=0.49\textwidth]{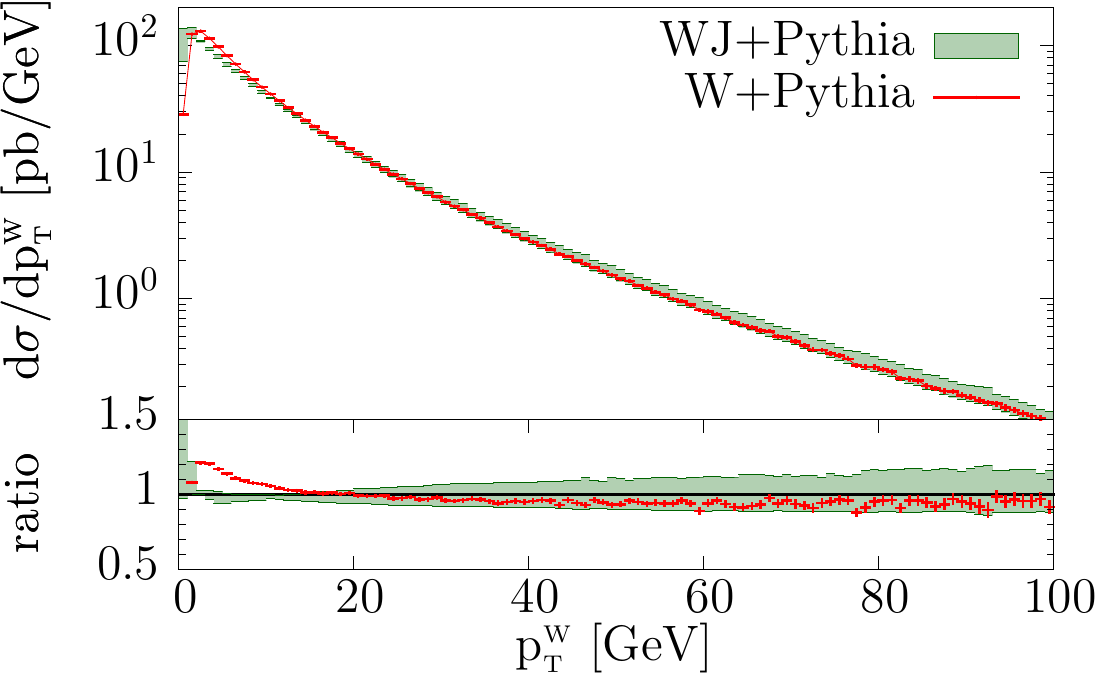}
\end{center}
\caption{Same as fig.~\ref{fig:Wpt} for a different $\pt^{\rm \sss W}$ range.}
\label{fig:Wptzm} 
\end{figure}
In figs.~\ref{fig:Wpt} and~\ref{fig:Wptzm} we present predictions for the
$W^-$ boson transverse-momentum spectrum. In this case we find noticeable
shape differences between the \W{} and \WJ{}+\MINLO{} distribution,
especially at low $\pt^{\rm \sss W}$. In particular, we observe that the
\WJ{}+\MINLO{} Sudakov form factor peaks at a lower value of $\pt^{\rm \sss
  W}$. These differences do not come as a surprise, since this distribution
is described only at LO by the \W{} generator, while the \WJ{}+\MINLO{}
description is NLO accurate. We also note that the \W{} uncertainty band is
small and uniform in the whole $\pt^{\rm \sss W}$ range.  This is a feature
of \POWHEG{} when no separation is performed between singular and regular
contributions to the cross section. In this case, the size of the scale
variation amounts to a factor $1+{\cal O}(\as^2)$, that is clearly too small
in the moderate to large $\pt^{\rm \sss W}$ region, where the \W{} generator
is only tree-level accurate. The error band given by the \WJ{}+\MINLO{}
generator is of an acceptable size at large transverse momenta, while it
seems to be excessively small in the very low transverse momentum region.

\begin{figure}[htb]
\begin{center}
\includegraphics[width=0.49\textwidth]{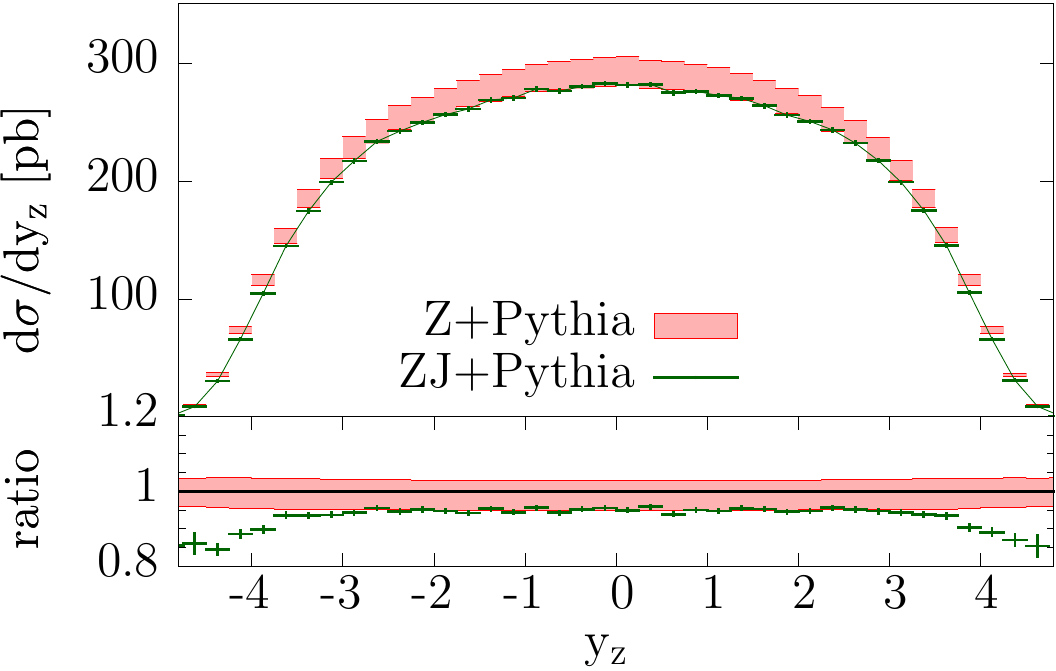}
\includegraphics[width=0.49\textwidth]{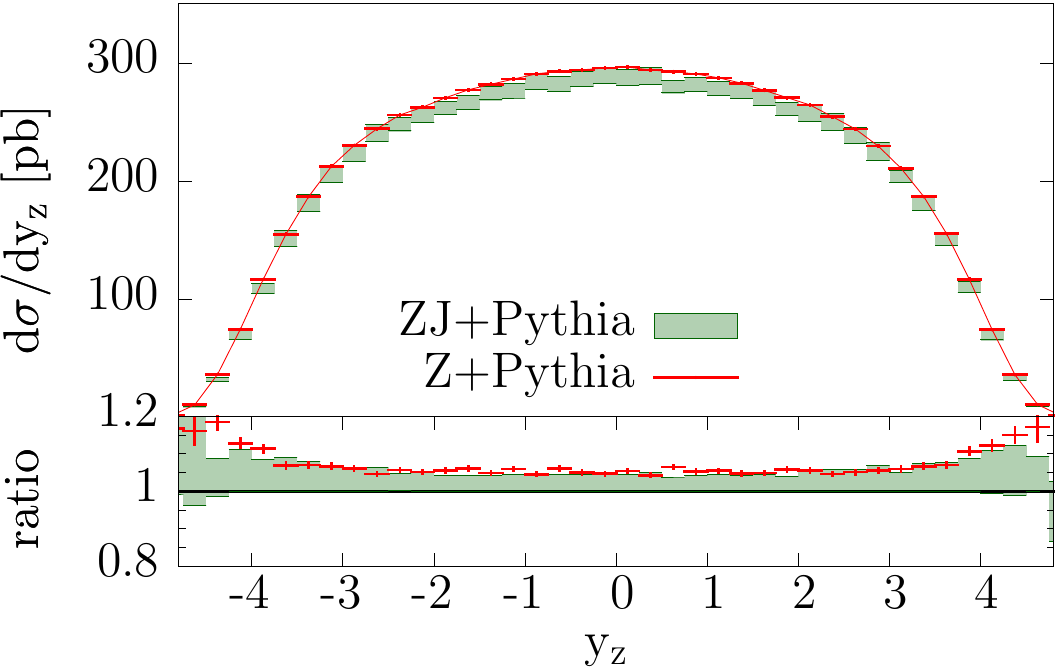}
\end{center}
\caption{Comparison between the \Z+\PYTHIA result and the
  \ZJ-\MiNLO+\PYTHIA{} result for the $Z$ rapidity distribution at the LHC at
  14~TeV. The bands are obtained as in fig.~\ref{fig:Wy}.}
\label{fig:Zy} 
\end{figure}
Finally, we discuss the case of $Z$ production at the 14~TeV LHC.  In
fig.~\ref{fig:Zy} we show the $Z$ boson rapidity distribution. As in the case
of $W$ production, we note that the \ZJ{}+\MINLO{} central value is lower
than the \Z{} one and that its uncertainty band is comparable at central
rapidities, widening in the forward-backward region. We also notice a slight
change of shape in the extreme rapidities, with the \Z{} result remaining
compatible with the \ZJ{}+\MINLO{} uncertainty band.

\begin{figure}[htb]
\begin{center}
\includegraphics[width=0.49\textwidth]{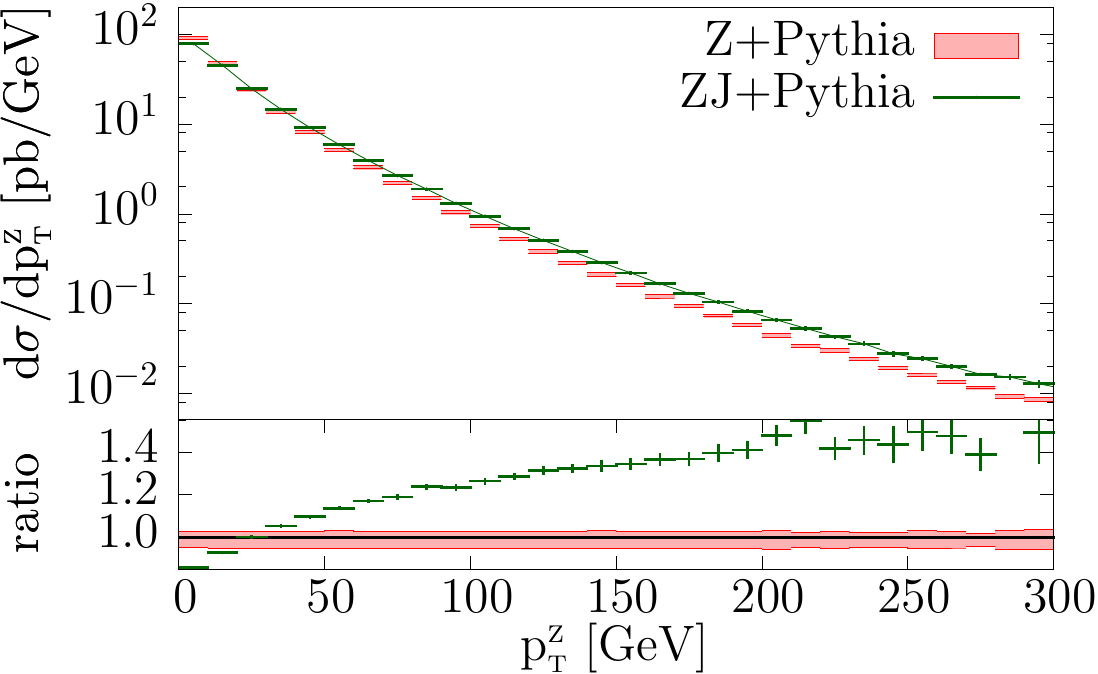}
\includegraphics[width=0.49\textwidth]{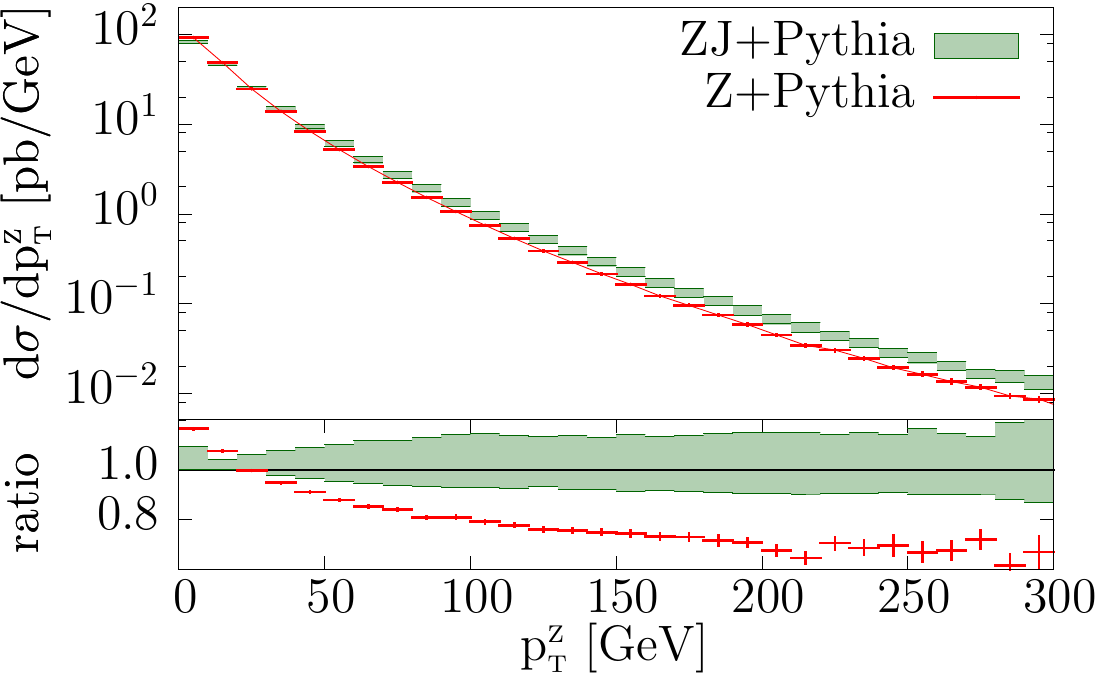}
\end{center}
\caption{Comparison between the \Z+\PYTHIA result and the
  \ZJ-\MiNLO+\PYTHIA{} result for the $Z$ transverse-momentum distribution at
  the LHC at 14~TeV. The bands are obtained as in fig.~\ref{fig:Wy}.}
\label{fig:Zpt} 
\end{figure}
\begin{figure}[htb]
\begin{center}
\includegraphics[width=0.49\textwidth]{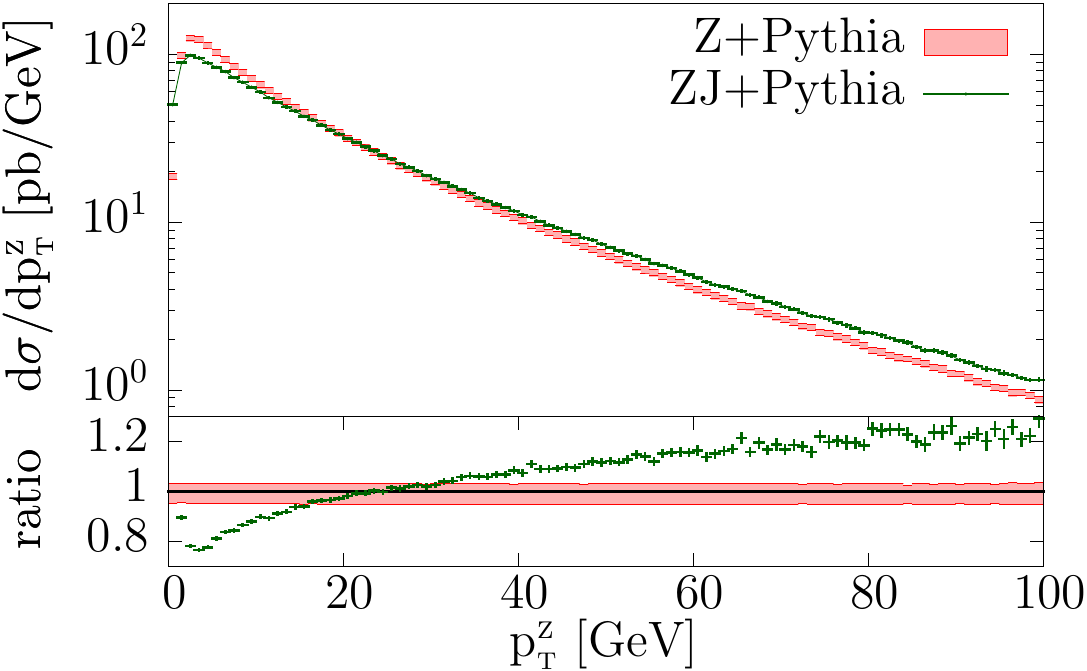}
\includegraphics[width=0.49\textwidth]{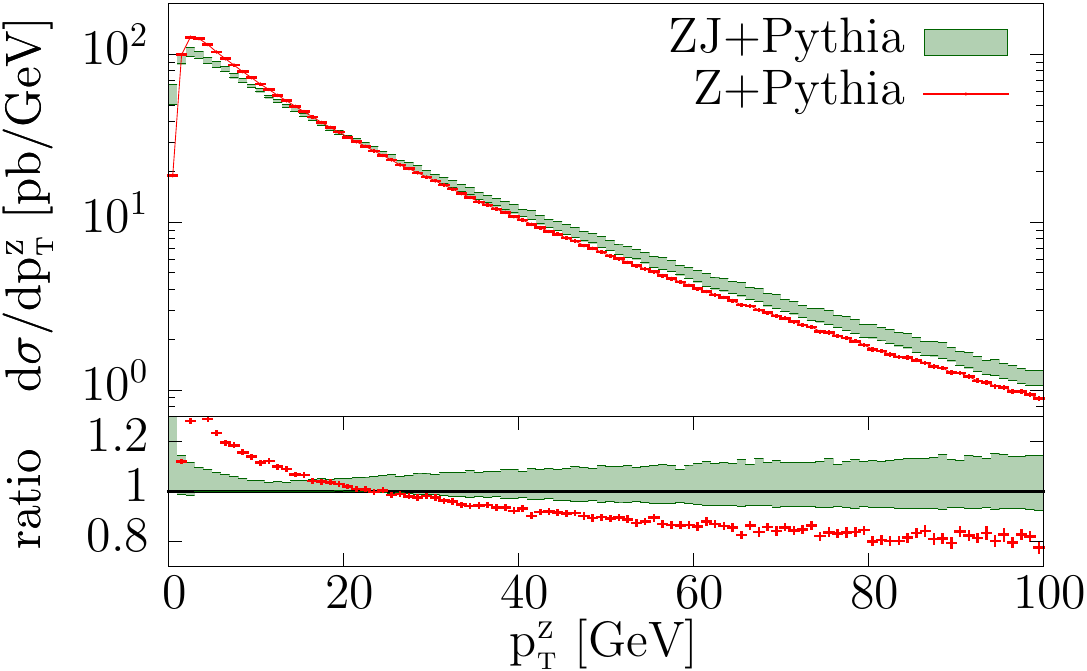}
\end{center}
\caption{Same as fig.~\ref{fig:Zpt} for a different $\pt^{\rm \sss Z}$ range.}
\label{fig:Zptzm} 
\end{figure}
In figs.~\ref{fig:Zpt} and \ref{fig:Zptzm} we show the transverse-momentum
distribution of the $Z$ boson.  We notice the same features already observed
in the $W$ case.  However we also observe now a considerable difference of
the \Z{} and \ZJ{}-\MiNLO{} distributions at large transverse momenta.
Following arguments in ref.~\cite{Rubin:2010xp} we suggest that this
difference arises due to NLO corrections in the \ZJ{}-\MiNLO{} generator which
are not present in the \Z{} one, for example those related to threshold
logarithms.

In conclusion, we have seen that in the case of vector-boson production ($Z$
and $W$), the agreement of the \BJ-\MiNLO{} and the \B{} generators is less
than perfect. We also have shown that there is not a close correspondence
between the scale variations in the two types of generators. We may conclude
from this observation that either the scale variation in the \MiNLO{}
generator is excessive, and should be limited (for example by considering
only 3-point scale variations), or that the scale uncertainty obtained in the
\W{} and \Z{} generators underestimates the true error. We also remark that
if we use the full 7-point scale variation for the \WJ{} and \ZJ{}
generators, the error bands for the transverse-momentum distributions do not
shrink as much at small transverse momenta, thus leading to better
compatibility with the \W{} and \Z{} generators distributions.  On the other
hand, by doing so, the error on the total cross section might be too
conservative. The problems with the application of our method to these
generators may be related to the fact that the bulk of the Sudakov region is
relatively low, the peak being below 5 GeV. We have ignored some problems
that may be relevant in this region. For example, the pdf evolution switches
from 5 to 4 and to 3 flavours in this region, while we have used constant
$n_f=5$ in the \BJ-\MiNLO{} generator.  Similarly, inaccuracies in the pdf
evolution in this region may also be relevant.  While we have checked that
our results are not affected in a relevant way by the infrared cutoffs of the
calculation, we believe that further work will be needed to better assess the
importance of low-$\pt$ Sudakov region.

On the contrary, in the Higgs boson case, we find from our study a fairly
good agreement between the \H{} and \HJ{}-\MiNLO{} generators, both for the
total cross sections and for the distributions. The scale variation bands are
also comparable, which confirms that our result yields an improved and more
accurate description of Higgs boson production, fulfilling also, in a
practical, way our goals.

\section{How to build an \NNLOPS{} generator with \MINLO{}}
\label{sec:NNLO}
We argue now that a generator built according to the \MiNLO{} prescription,
improved with the findings of the present work, can be easily turned into an
\NNLOPS{} generator. For the sake of simplicity, we consider the Higgs boson
generator.  We have concluded that such a simulation achieves ${\cal
  O}(\as^4)$ accuracy for all distributions involving at least one jet and
${\cal O}(\as^3)$ accuracy for inclusive distributions. We denote by
\begin{equation}
\left(\frac{\mathd \sigma}{\mathd y }\right)_{\mbox{\HJ{}}}
\end{equation}
the inclusive Higgs boson rapidity distribution obtained with this event
generator.  We denote by
\begin{equation}
\left(\frac{\mathd \sigma}{\mathd y }\right)_{\mbox{\scriptsize{NNLO}}}
\end{equation}
the inclusive Higgs boson rapidity distribution calculated with a fixed NNLO
calculation. We claim that by reweighting the \HJ-\MiNLO{} output with the
weight factor
\begin{equation}
\frac{\left(\frac{\mathd \sigma}{\mathd y }\right)_{\mbox{\scriptsize{NNLO}}}}
{\left(\frac{\mathd \sigma}{\mathd y }\right)_{\mbox{\HJ{}}}}
\end{equation}
we achieve full NNLO accuracy for our generator. The proof is simple. The above
ratio has a formal expansion
\begin{equation}
\frac{\left(\frac{\mathd \sigma}{\mathd y }\right)_{\mbox{\scriptsize{NNLO}}}}
{\left(\frac{\mathd \sigma}{\mathd y }\right)_{\mbox{\HJ{}}}}=
\frac{c_2\as^2+c_3\as^3+c_4\as^4}{c_2\as^2+c_3\as^3+d_4\as^4}\approx
1+ \frac{c_4-d_4}{c_2}\as^2+{\cal O}(\as^3).
\end{equation}
Notice that the numerator and denominator agree at ${\cal O}(\as^3)$ since
the \HJ{}-\MiNLO{} generator achieves ${\cal O}(\as^3)$ accuracy for
inclusive observables.  Thus, the reweighting factor does not spoil the
$\as^4$ accuracy of the \HJ-\MiNLO{} generator in the one-jet region. In this
region, in fact, the dominant contributions to the \HJ-\MiNLO{} generator are
of order $\as^3$, and the reweighting generates extra contributions of order
$\as^5$, that are beyond the nominal accuracy. On the other hand, the
inclusive distributions are reweighted to achieve $\as^4$ accuracy, so that
the generator has indeed $\as^4$ accuracy in the whole phase space.

Variants of these schemes are also possible.  Rather than reweighting using
the full rapidity distribution, one can split the cross section as
\begin{equation}
\frac{\mathd^2 \sigma}{\mathd {\qT} \mathd y } = 
\frac{\mathd^2 \sigma}{\mathd {\qT} \mathd y } \, h\(\qT\)+
\frac{\mathd^2 \sigma}{\mathd {\qT} \mathd y } \left[1-h\(\qT\)\right]\,, 
\end{equation}
where $h$ is a smooth positive function such that $h(\qT)\to1$ as $\qT\to 0$
and $h(\qT)\to0$ for $\qT\gg \MH$, such as
\begin{equation}
h(q)=\frac{c\, \MH^\alpha}{c\, \MH^\alpha+ q^\alpha},
\end{equation}
with $\alpha\ge 1$ and $c$ a constant of order 1.  One can then reweight the cross
section as
\begin{equation}
\frac{\mathd^2 \sigma}{\mathd {\qT} \mathd y } \,h\(\qT\) W\(y \) +
\frac{\mathd^2 \sigma}{\mathd {\qT} \mathd y } \left[1-h\(\qT\)\right]\,, 
\end{equation}
with 
\begin{equation}
W(y ) = \frac{\displaystyle \int\mathd\qT \, h\(\qT\) \left(\frac{\mathd^2
    \sigma}{\mathd\qT \mathd y
  }\right)_{\mbox{\scriptsize{NNLO}}}}{\displaystyle\int\mathd\qT\, h\(\qT\)
  \left(\frac{\mathd^2 \sigma}{\mathd \qT \mathd y }\right)_{\HJ{}}}\,.
\end{equation}
In this way, for $\qT\gg \MH$ the effect of the reweighting vanishes.

We notice that this \NNLOPS{} generator would be NNLO accurate in the same
sense in which the current \MCatNLO{} or \POWHEG{} type generators are NLO
accurate, i.e.~integrated quantities achieve NLO/NNLO accuracy, while
LL/NLL/NNLL accuracy is achieved in the Sudakov region, depending upon the
accuracy of the implementation of the Sudakov form factors.

We postpone a phenomenological study including reweighting to a future
publication.

There are a number of more complex processes to which this procedure can be
generalized, typically processes where the heavy particle decays, or
processes involving pair production of massive colourless objects. In this
case, the NNLO reweighting must be performed as a function of more
variables. One would typically use the rapidity of the heavy system, plus the
kinematical variables describing its internal structure.  For example, one
can go to the longitudinal rest frame of the heavy system with a Lorentz
boost, perform a transverse boost of the system such that its transverse
momentum vanishes, and use variables that describe the kinematics of the
heavy system in this frame.

\section{\MiNLO{} merging for more complex processes}
\label{sec:Prospects}
In the present work, we have dealt with relatively simple processes, i.e.~a
colour-neutral massive particle in association with one jet. The results of
ref.~\cite{Hamilton:2012np} strongly suggest that our procedure may be
generalized to higher jet multiplicities.  Since, at the level of one
associated jet, we had to improve the \MINLO{} prescription with the
inclusion of certain NNLL terms in the Sudakov form factor, it is clear that,
in general, we need to find a similar improvement for the more general case.

A first question that we would like to consider is whether our \MiNLO{}
procedure downgraded to the LO level (that is to say, to the CKKW
procedure~\cite{Catani:2001cc} as applied to the inclusive sample), already
achieves our goal at LO accuracy, that is to say, it is such that by
integrating the softest emission one gets a LO accurate matrix element for
one less emission. It is easy to convince ourselves that as soon as we deal
with matrix elements involving more than four coloured particles (including
the softest emission), this is not the case.  In fact, after the first
clustering, that in our procedure simply sets the $Q_0$ scale, we are left
with four or more coloured partons. Soft gluon resummation, in this case,
also involve soft, non-collinear terms that arise from interference of the
emission from the coloured external lines~\cite{Kidonakis:1996aq,
  Kidonakis:1997gm, Bonciani:2003nt}. These terms are of NLL accuracy, and,
according to our counting, they can contribute terms of relative order
$\sqrt{\as}$. By not including them, we thus introduce an error of this
magnitude, while LO accuracy requires the neglected terms to be of order
$\as$. Notice that this problem does not manifest itself in the \HJJ{} case,
since precisely four coloured partons are present here (two in the initial
state and two in the final state).  In processes like $t\bar{t}$ production
in association with one jet, such terms would have to be accounted for. On
the other hand, it is possible to compute these terms using standard
resummation techniques.
It is thus conceivable that the LO \MiNLO{} formula can also be improved
including these interference terms. But we also stress that the CKKW
procedure, as is, does not satisfy our requirement at leading order.

In the present work, we have used as a clustering variable the transverse
momentum of the boson, since it is a simple variable and the corresponding
resummation formulae are well known.  As an alternative, we could have used
the hardest jet transverse momentum, taking $B_2$ from
ref.~\cite{Banfi:2012jm}. If this method is to be extended to processes with
more than one radiated parton, it is clear that other clustering variables
should be chosen, likely with good resummation properties. One should then
seek either an NNLL extension of the CKKW procedure, or, construct a product
of standard soft resummation factors, accurate at the NLL level, modifying
the Sudakov form factor in each one to include the $B_2$ term relevant to the
associated configuration of clustered particles. Notice that within our
method it was never necessary to know explicitly the hard resummation
correction that is usually included in NLL and NNLL resummation formulae,
since the NLO matrix elements intrinsically provide this.

\section{Conclusions}
\label{sec:conclu}
In the present work, we have illustrated a method for constructing \NLOPS{}
generators for the production of a heavy system accompanied by a radiated
parton, such that, when integrating over the parton's phase space, one
recovers the accuracy of a corresponding \NLOPS{} generator for the
production of the heavy system alone. In essence, in our method, we start
from the \MiNLO{} prescription of ref.~\cite{Hamilton:2012np}, and we look
explicitly for the places in which it needs to be modified in order to
maintain NLO accuracy on integrating out all radiation. We have found that
the inclusion of the $A_2$ and the $B_2$ term in the resummation formula is
enough to achieve this goal. Since we do not know the $B_2$ term for the type
of clustering that we performed in ref.~\cite{Hamilton:2012np}, we have
modified the prescription so that the transverse momentum of the boson is
used instead. More specifically, the prescription is applicable to any
clustering scheme in which the last step is performed using the transverse
momentum of the boson as the clustering scale.

We have not explored, at this moment, any issues related to the logarithmic
accuracy of our approach. In other words, the logarithmic accuracy should be
at the same level as that of the \NLOPS{} generators we are referring to. We
postpone to future studies, a more accurate assessment of, and possible
improvements to, the precision of the resummation.

We have tested our method in the framework of \HWZ{} production. We find that
the method performs remarkably well. We have also found that the usual
scale-variation method used in order to determine uncertainties may be
deceiving in our case, especially for processes like \WZ{} production, where,
at the Born level, there is no renormalization scale dependence.  We track
this problem to the fact that, in the corresponding \BJ{} calculation, a
renormalization scale variation is already possible at the Born level.

We point out that, using our \BJ{}-\MiNLO{} generators, it is actually
possible to construct an \NNLOPS{} generator, simply by reweighting the
transverse-momentum integral of the cross section to the one computed at the
NNLO level. We postpone a phenomenological study of this method to a future
publication.

While this work was under completion, a publication has appeared that has
some points in common with our work~\cite{Alioli:2012fc}. This work focuses
on the accuracy of the matching conditions, and sets up a framework, using
NNLL accurate resummation of soft-parton emission, such that one has no loss
of NLO accuracy when matching.  Also our approach is motivated by the
requirement of preserving the NLO accuracy. It is more focussed, however,
upon finding the minimal modification to the soft-parton resummation such
that no matching is needed at all, and, in fact, we identify precisely where
do we need to improve the resummation formula in order to achieve our goal.
In this way, we find that its implementation in the \BJ{} case requires only
a minimal modification to the \MiNLO{} procedure, that, by itself, is quite
simple.

\section*{Acknowledgments}
\label{sec:Acknowledgments}
We thank Gavin Salam and Massimiliano Grazzini for useful discussions. 
G.Z.\ is supported by the British Science and Technology Facilities Council.
G.Z., P.N. and C.O. acknowledge the support also of grant PITN-GA-2010-264564
from the European Commission.

\appendix

\section{The NNLL resummed differential cross section}
\label{app:qtresum} 
The NNLL differential cross section for the production of a colourless
system, denoted by $B$, of virtuality $Q^2$ (when dealing with a single
vector boson, we assume that $Q$ is equal to the vector boson mass $M$), in
the partonic scattering $i + j \to B$, is given by
\begin{equation}
\label{eq:dsig_dqtdy}
  \frac{\mathd \sigma}{\mathd \qt^2 \mathd \yB } =  \sigma_0\! 
  \int\! \mathd^2 b \, e^{i \vec{q}_{\sss\rm T} \cdot \vec{b}}  \left[ C_{ia} \otimes
  f_{a / A} \right]\! \left( x_A, \frac{c_1}{b} \right)  \left[ C_{ jb} \otimes
  f_{b / B} \right] \!\left( x_{\Beta}, \frac{c_1}{b} \right) \times \exp{\cal S} \left(
  c_2 Q, \frac{c_1}{b} \right),
\end{equation}
where we have used the following definition of convolution
\begin{equation}
\left[ C_{ij} \otimes f_{j / J} \right]\! \left( x, q \right)
\equiv 
\int_{x}^1 \frac{\mathd \xi}{\xi} \, f_{i / J} \(\xi, q\) 
C_{ij} \(\as\(q^2 \) ,\frac{x}{\xi}\).
\end{equation}
The $\sigma_0$ factor is such that, at leading order,
\begin{equation}
 \( \frac{\mathd \sigma}{\mathd \yB}\)_{\rm LO} = f_{i / A} \left( x_A, \mu_F
 \right) f_{j / \Beta} \left( x_{\Beta}, \mu_F \right)
   \sigma_0\, ,
\end{equation}
and the rapidity of the $B$ system is given by
\begin{equation}
   \yB = \frac{1}{2} \log \frac{x_A}{x_B} \,.
\end{equation}
The Sudakov form factor $\mathcal{S} \left( q', q \right)$ is defined
by~\footnote{Note that this Sudakov is the square of the Sudakov form factor
  $\Delta_{q/g}$ considered in Sec.~\ref{sec:accuracy} (see
  eq.~\ref{eq:calSdef}). Therefore the coefficients $A_i$ and $B_i$ here are
  twice those given in eqs.~(\ref{eq:AiBiq}) and~(\ref{eq:AiBig}).}
\begin{equation}
  \mathcal{S} \left( q', q \right) = - \int_{q^2}^{q'^2} \frac{\mathd
  \mu^2}{\mu^2} \left[ A \left( \as \left( \mu^2 \right) \right) \log
  \frac{q'^2}{\mu^2} + B \left( \as \left( \mu^2 \right) \right) \right],
\end{equation}
and the $C_{ij}$ coefficients have the following perturbative expansion
\begin{equation}
  C_{i j} \equiv C_{i j} \left( \as, z \right) =
  \delta_{i j} \delta \left( 1 - z \right) + \as \, C_{i j}^{(1)} \left( z \right)+
   \ldots
\end{equation}
The $C_{ij}$ functions can thus be absorbed into a redefinition of the
pdfs. We will thus assume now that our pdfs include this factor, and only
at the end we will reinstate it.

Equation~(\ref{eq:dsig_dqtdy}) is easily manipulated by going to Mellin
space. However, we do not want to lose the information on the rapidity $\yB$
of the $B$ system. We then define the following Fourier-Mellin transform
\begin{equation}
  \sigma_N  \equiv \int \mathd\tau \, \mathd \yB \, \tau^{ \alpha- 1}
  \, e^{i \, 2\,\beta \,\yB} \frac{\mathd
    \sigma}{\mathd \qT^2 \mathd \yB}
\end{equation}
where
\begin{equation}
  \tau = x_A x_B\,,
\end{equation}
so that $S = Q^2 / \tau$, where $S$ is the hadronic center-of-mass energy.
Notice that the $\tau$ integration is performed by considering that $S$ goes
from $Q^2$ to infinity. We can then introduce a complex number $N$ such that
$N=\alpha + i\, \beta$, so that we can write
\begin{eqnarray}
\label{eq:sigN}
  \sigma_N & =
& \int_0^1 \mathd x_A \, \mathd x_B \, \tau^{\tmop{Re} \left[ N
  \right] - 1} \, e^{i \, 2 \, \tmop{Im} \left[ N \right] \, \yB}  \frac{\mathd
  \sigma}{\mathd \qT^2 \mathd \yB} \nonumber\\
  & = & \int_0^1 \mathd x_A \, \mathd x_B \,  \left( x_A x_B \right)^{\tmop{Re} \left[
  N \right] - 1}  \left( \frac{x_A}{x_B} \right)^{i \tmop{Im} \left[ N
  \right]}  \frac{\mathd \sigma}{\mathd \qT^2 \mathd \yB} \nonumber\\
  & =
 & \int_0^1 \frac{\mathd x_A \,\mathd x_B}{x_A\, x_B} \, x_A^N \, x_B^{N^{\ast}} 
  \frac{\mathd \sigma}{\mathd \qT^2 \mathd \yB} \nonumber\\
  & = & \sigma_0  \int \mathd^2 b \, e^{i \vec{q}_{\sss\rm T} \cdot \vec{b}}\,
  f_{a /A,N}\left( \frac{c_1}{b} \right) \, f_{b /B, N^{\ast}} \left(  \frac{c_1}{b} \right)
  \, \exp \mathcal{S} \left( c_2  Q, \frac{c_1}{b} \right),
\end{eqnarray}
with 
\begin{equation}
f_{a /A,N}\left( \frac{c_1}{b} \right) = 
\int_0^1 {\mathd x_A}  \, x_A^{N-1} f_{a/A}(x_A,\frac{c_1}{b})\,.
\end{equation}
By solving the DGLAP equations for the pdfs to the required logarithmic
accuracy, and evolving the moments of the $C_{ij}$ coefficients too,  we can write
\begin{eqnarray}
  f_{a /A, N} \left(\frac{c_1}{b} \right) \, f_{b /B, N^{\ast}} \left(
  \frac{c_1}{b} \right) &=& f_{a / A, N}\left( c_2 Q \right)\, f_{b / B,
    N^{\ast}} \left( c_2 Q \right) \, \exp \left\{ \tmop{Re} \left[ {\cal G}_N
      \left( c_2 Q, \frac{c_1}{b} \right) \right] \right\}
\nonumber \\
&\equiv& F_N(c_2 Q) \, \exp \left\{ \tmop{Re} \left[ {\cal G}_N
      \left( c_2 Q, \frac{c_1}{b} \right) \right] \right\},
\end{eqnarray}
with (see eq.~(2.22) of ref.~\cite{Frixione:1998dw})
\begin{equation}
\label{eq:G_N_Q}
  \mathcal{G_{}}_N \left( c_2 Q, \frac{c_1}{b} \right) = - 2 \int_{c_1^2 /
    b^2}^{c_2^2 Q^2} \frac{\mathd \mu^2}{\mu^2} \, \gamma_N \left( \as \left(
  \mu^2 \right) \right).
\end{equation}
We define
\begin{eqnarray}
  \as & = & \as \left( c_2^2 Q^2 \right)\,, \\
  y & = & - \as\, b_0 \log \frac{c_1^2}{c_2^2 \, b^2\,  Q^2} \,,
\end{eqnarray}
and, using the solution the renormalization group equation for $\as$,
\begin{equation}
\frac{\mathd \as}{\mathd \log \mu^2}=-b_0 \as^2-b_1\as^3 -b_2 \as^4 +\ldots\,,
\end{equation}
we write
\begin{equation}
  \mathcal{S} \left( c_2 Q, \frac{c_1}{b} \right) \equiv \mathcal{S} \left(
  \as, y \right) = \frac{1}{\as} f_0 \left( y \right) + f_1 \left( y
  \right) + \as f_2 \left( y \right)+\ldots,
\end{equation}
where
\begin{eqnarray}
  f_0 \left( y \right) & = & \frac{A_1}{b_0^2} \left[ y + \log \left( 1 - y
  \right) \right], \nonumber\\
  f_1 \left( y \right) & = & \frac{A_1 b_1}{b_0^3}  \left[ \frac{1}{2} \log^2
  \left( 1 - y \right) + \frac{y}{1 - y} + \frac{\log \left( 1 - y \right)}{1
  - y} \right] \nonumber\\
  &  &{}- \frac{A_2}{b_0^2}  \left[ \log \left( 1 - y \right) + \frac{y}{1 - y}
  \right] + \frac{B_1}{b_0} \log \left( 1 - y \right), \nonumber\\
  f_2 \left( y \right) & = & - \frac{B_2}{b_0} \frac{y}{\left( 1 - y \right)}
  + \frac{B_1 b_1}{b_0^2}  \frac{y + \log \left( 1 - y \right)}{1 - y} -
  \frac{A_3}{2 b_0^2}  \frac{y^2}{\left( 1 - y \right)^2} \nonumber\\
  &  &{}+ \frac{A_2 b_1}{2 b_0^3}  \frac{3 y^2 - 2 y + \left( 4 y - 2 \right)
  \log \left( 1 - y \right)}{\left( 1 - y \right)^2} + \frac{A_1}{2 b_0^4} 
  \frac{1}{\left( 1 - y \right)^2}  \Big\{ b_1^2 \left( 1 - 2 y \right) \log^2
  \left( 1 - y \right)  \nonumber\\
  & & {}+2 \left[ b_0 b_2 \left( 1 - y \right)^2 + b_1^2 y \left( 1 - y
  \right) \right] \log \left( 1 - y \right) - 3 b_0 b_2 y^2 + b_1^2 y^2 + 2
  b_0 b_2 y \Big\}. 
\end{eqnarray}
Notice that the coefficients $A_2$, $B_1$ and $B_2$ do depend explicitly upon
$c_1$ and $c_2$ (see ref.~\cite{Collins:1984kg}).  The values $c_2 = 1$ and
$c_1 = 2 e^{- \gamma_E}$ give the coefficients that are usually reported in
the literature.

Similarly, we can expand $\mathcal{G}_N \left( c_2 Q, c_1 / b
\right)$, using 
\begin{equation}
\gamma_N(\as) = \as \gamma_{1,N}+\as^2 \gamma_{2,N}\,,
\end{equation}
we can write
\begin{equation}
  \mathcal{G}_N \left( c_2 Q, \frac{c_1}{b} \right) \equiv \mathcal{G}_N
  \left( \as, y \right) = g_{1, N} \left( y \right) + \as \, g_{2, N}
  \left( y \right),
\end{equation}
with
\begin{eqnarray}
\label{eq:g1n}
  g_{1, N} \left( y \right) &=& \frac{\gamma_{1, N}}{b_0} \log \left( 1 - y
  \right) , \hspace{2em} 
\\
g_{2, N} \left( y \right) &=& \frac{2}{b_0^2 \left( 1 - y \right)} \left[
  \gamma_{1, N} b_1 \log \left( 1 - y \right) - \left( b_0 \gamma_{2, N} -
  b_1 \gamma_{1, N} \right) y\right]\, .
\end{eqnarray}
We first perform the angular integration in eq.~(\ref{eq:sigN}), we change
the integration variable to $\hat{b} = b \qt$ and we integrate by part, to get
(see ref.~\cite{Frixione:1998dw} for more details) 
\begin{equation}
  \frac{\mathd \sigma_N}{\mathd \qt^2} = F_N \left( c_2 Q \right) \,
  \frac{\mathd}{\mathd \qt^2} \int_0^{\infty} \!\mathd \hat{b} \, J_1 (
  \hat{b} ) \,\exp \mathcal{S} \left( c_2 Q, c_1 \frac{\qt}{\hat{b}}
  \right) \, \exp \mathcal{G}_N \left( c_2 Q, c_1 \frac{\qt}{\hat{b}} \right),
\end{equation}
at NNLL accuracy.
 We then write
\begin{equation}
  y = y_0 + \as b_0 l_b,
\end{equation}
with
\begin{equation}
  y_0 = - \as b_0 \log \frac{\qt^2}{c_2^2 Q^2}\,, \qquad \qquad
 l_b =  \log \frac{\hat{b}^2}{c_1^2} \,.
\end{equation}
Now we have to consider $l_b$ as not being parametrically large, while $y_0$
is the variable that becomes of order 1 in the small $\qt$ limit. We thus
expand $\mathcal{S}$ and $\mathcal{G}_N$ at NNLL order
\begin{eqnarray}
  \mathcal{S_{}}_N \left( \as, y \right) & = & \mathcal{S_{}}_N \left( \as,
  y_0 \right) + f'_0 \left( y_0 \right) b_0 l_b + \frac{\as}{2} f''_0 \left(
  y_0 \right) b_0^2 l_b^2 + \as f'_1 \left( y_0 \right) b_0 l_b\,,
  \\ 
\label{eq:S_N}
\mathcal{G}_N \left( \as, y \right) & = & \mathcal{G_{}}_N \left(
  \as, y_0 \right) + \as g'_{1, N} \left( y_0 \right) b_0 l_b \,.
\label{eq:G_N}
\end{eqnarray}
We get
\begin{eqnarray}
  \frac{\mathd \sigma_N}{\mathd \qt^2} & = & F_N \left( c_2 Q \right) 
  \frac{\mathd}{\mathd \qt^2} \exp \mathcal{S} \left( \as, y_0
  \right) \, \exp \big\{ \tmop{Re} \left[  \mathcal{G}_N \left( \as, y_0
  \right) \right] \big\} \int_0^{\infty} \mathd \hat{b} \, J_1 ( \hat{b}) 
\nonumber\\
  & &{}\times  \exp \left[ f'_0 \left( y_0 \right) b_0 l_b + \as f'_1
  \left( y_0 \right) b_0 l_b + \as \tmop{Re} \left[ g'_{1, N} \left( y_0
  \right) \right] b_0 l_b + \frac{\as}{2} f''_0 \left( y_0 \right) b_0^2
  l_b^2 \right]\!. \phantom{aaaaa}
\end{eqnarray}
We now work out the integral
\begin{eqnarray}
  I & = & \int_0^{\infty} \mathd \hat{b} J_1(\hat{b}) \exp
  \left\{ f'_0 \left( y_0 \right) b_0 l_b + \as f'_1 \left( y_0 \right)
  b_0 l_b + \as \tmop{Re} \left[ g'_{1, N} \left( y_0 \right) \right] b_0 l_b + \frac{\as}{2}
  f''_0 \left( y_0 \right) b_0^2 l_b^2 \right\} 
\nonumber\\
  & = & \int_0^{\infty} \mathd \hat{b} \, J_1(\hat{b})  \left\{ 1 +
  \as b_0 \Big[ f'_1 \left( y_0 \right) + \tmop{Re} \left[g'_{1, N} \left( y_0 \right)\right]
  \Big] 2 \frac{\partial}{\partial h} + \frac{\as}{2} f''_0 \left( y_0
  \right) b_0^2 4 \frac{\partial}{\partial h^2} \right\}  \hat{b}^h  \left(
  \frac{1}{c_1} \right)^h 
\nonumber\\
  & = & \left\{ 1 + 2 \as b_0 \Big[ f'_1 \left( y_0 \right) + \tmop{Re}
    \left[ g'_{1, N}  \left( y_0 \right)\right] \Big] \frac{\partial}{\partial h} + 2 \as f''_0
  \left( y_0 \right) b_0^2  \frac{\partial}{\partial h^2} \right\}  \left(
  \frac{2}{c_1} \right)^h \frac{\Gamma \left( 1 + h / 2 \right)}{\Gamma \left(
  1 - h / 2 \right)}, 
\nonumber\\
 \label{eq:I2}
\end{eqnarray}
where
\begin{equation}
  h = 2 f'_0 \left( y_0 \right) b_0 = - \frac{2 A_1}{b_0} \frac{y_0}{1 - y_0}  .
\end{equation}
We then obtain
\begin{eqnarray}
\label{eq:I}
  I & = &\left\{ 1 + \as b_0 \Big[ f'_1 \left( y_0 \right) + 
  \tmop{Re} \left[  g'_{1, N} \left( y_0 \right) \right] \Big] \left[ \psi_0 \left( 1 + \frac{h}{2}
  \right) + \psi_0 \left( 1 - \frac{h}{2} \right) + 2 \gamma_E
  \right]\right. 
\nonumber\\
 && {}+ \frac{\as}{2} f''_0 \left( y_0 \right) b_0  \left[ \psi_1 \left( 1 +
  \frac{h}{2} \right) + \psi_0^2 \left( 1 + \frac{h}{2} \right) + 2 \psi_0
  \left( 1 - \frac{h}{2} \right) \psi_0 \left( 1 + \frac{h}{2} \right)
  \right.
\nonumber\\
&& {} +4 \gamma_E \psi_0 \left( 1 + \frac{h}{2} \right) - \psi_1 \left( 1 -
  \frac{h}{2} \right) + \psi_0^2 \left( 1 - \frac{h}{2} \right)  
\nonumber\\
&& {}+\left. \left.  4 \gamma_E \psi_0 \left( 1 - \frac{h}{2} \right) + 4 \gamma_E^2 \right]
  \right\} e^{\gamma_E h}  \frac{\Gamma \left( 1 + \frac{h}{2} \right)}{\Gamma
  \left( 1 - \frac{h}{2} \right)}\,,
\end{eqnarray}
where
\begin{equation}
\psi_0(z) = \frac{\mathd}{\mathd z} \Gamma(z)\,.
\end{equation}
Notice that the $N$-dependent term, that prevents rewriting the luminosity in
$x$ space, depends upon $N$ through the factor $g'_{1, N}$. Thus, the
$N$-dependent term can be factorized with the required accuracy as
\begin{eqnarray}
& & \exp \big\{ \tmop{Re} \left[ \mathcal{G}_N \(\as,y_0\) \right]\big\}\, 
\exp \big\{ \as \,b_0\, K \, \tmop{Re} \left[ g'_{1, N} \left( y_0 \right) \right] \big\}
\nonumber\\
  & & \hspace{0.5cm} = \exp \frac{\mathcal{G}_N \(\as,y_0\)}{2} \, 
\exp \frac{ \as \, b_0\, K \, g'_{1,N}\( y_0\)}{2}
\exp \frac{\mathcal{G}_{N^*} \(\as,y_0\)}{2} \, 
\exp \frac{ \as \, b_0\, K \, g'_{1,N^*}\( y_0\)}{2}
\nonumber\\
  & & \hspace{0.5cm} = \exp \left\{ \half \left[
\mathcal{G}_N \(\as,y_0\) +  \as \, \, g'_{1,N}\( y_0\) b_0 \log\(\exp K\) \right]\right\}
\nonumber\\
&& \hspace{0.5cm}  \times 
\exp \left\{ \half \left[
\mathcal{G}_{N^*} \(\as,y_0\) +  \as \, \, g'_{1,N^*}\( y_0\) b_0 \log\(\exp K\) \right]\right\}
\nonumber\\
  &  &  \hspace{0.5cm} =  \exp\left\{ \half
\mathcal{G}_N \!\left( c_2 Q, \qt \exp \frac{K}{2}\right) \right\} \,  
 \exp\left\{ \half
\mathcal{G}_{N^*} \!\left( c_2 Q, \qt \exp \frac{K}{2}\right) \right\} \,, 
\end{eqnarray}
where
\begin{equation}
   K = \psi_0 \left( 1 + \frac{h}{2}
  \right) + \psi_0 \left( 1 - \frac{h}{2} \right) + 2 \gamma_E  .
\end{equation}
and we have used eqs.~(\ref{eq:G_N}) and the definition of $\mathcal{G}_N$
in eq.~(\ref{eq:G_N_Q}).

We can now perform the inverse Mellin/Fourier transform back to $x$ space,
obtaining
\begin{equation}
  \frac{\mathd \sigma_N}{\mathd \qt^2 \mathd \yB} = \frac{\mathd}{\mathd
  \qt^2}  \left[ C_{i a} \otimes f_{a / A} \right]\! \left( x_A, \qT
  \exp \frac{K}{2} \right)  \left[ C_{j b} \otimes f_{b / B} \right]\! \left(
  x_{B}, \qT \exp \frac{K}{2} \right) \exp \mathcal{S} \left( \as,
  y_0 \right) \,  I_0\,,
\end{equation}
where $I_0$ is equal to eq.~(\ref{eq:I}) with the $g'_{1, N}$ term deleted
\begin{eqnarray}
  I_0 & = &\left\{ 1 + \as b_0  f'_1 \left( y_0 \right) \left[ \psi_0 \left( 1 + \frac{h}{2}
  \right) + \psi_0 \left( 1 - \frac{h}{2} \right) + 2 \gamma_E
  \right]\right. 
\nonumber\\
 && + \frac{\as}{2} f''_0 \left( y_0 \right) b_0  \left[ \psi_1 \left( 1 +
  \frac{h}{2} \right) + \psi_0^2 \left( 1 + \frac{h}{2} \right) + 2 \psi_0
  \left( 1 - \frac{h}{2} \right) \psi_0 \left( 1 + \frac{h}{2} \right)
  \right.
\nonumber\\
&&  +4 \gamma_E \psi_0 \left( 1 + \frac{h}{2} \right) - \psi_1 \left( 1 -
  \frac{h}{2} \right) + \psi_0^2 \left( 1 - \frac{h}{2} \right)  
\nonumber\\
&&+\left. \left.  4 \gamma_E \psi_0 \left( 1 - \frac{h}{2} \right) + 4 \gamma_E^2 \right]
  \right\} e^{\gamma_E h}  \frac{\Gamma \left( 1 + \frac{h}{2} \right)}{\Gamma
  \left( 1 - \frac{h}{2} \right)} . 
\end{eqnarray}
At the moment we are not interested in developing this expression, that is
interesting of its own, any further. For our purpose, we just need to expand
$I_0$ to get
\begin{eqnarray}
  I_0 & = & \left[ 1 + 2 \as b_0 f'_1 \left( y_0 \right)
  \frac{\partial}{\partial h} + 2 \as f''_0 \left( y_0 \right) b_0^2 
  \frac{\partial}{\partial h^2} \right]  \left( \frac{2}{c_1} \right)^h
  \frac{\Gamma \left( 1 + h / 2 \right)}{\Gamma \left( 1 - h / 2 \right)}
\nonumber\\
  &\approx& 
  \left[ 1 + 2 \as b_0 f'_1 \left( y_0 \right) \frac{\partial}{\partial
  h} + 2 \as f''_0 \left( y_0 \right) b_0^2  \frac{\partial}{\partial
  h^2} \right] \left[ 1 - \frac{\zeta_3 h^3}{12} + \mathcal{O} \left( h^5
  \right) \right] 
\nonumber\\
&\approx&
  1 - \frac{\zeta_3 h^3}{12} - \frac{\as}{2} b_0 f'_1 \left( y_0 \right)
  \zeta_3 h^2 - \as f''_0 \left( y_0 \right) b_0^2 \zeta_3 h + \ldots 
  \label{eq:Iexpand}
\end{eqnarray}
Thus, the factor $I_0$ corrects the formula in ref.~\cite{Dokshitzer:1978dr}
by terms of the following order
\begin{eqnarray}
  h^3  \approx  \as^3 L^3  , \hspace{2em} &&
3  \mbox{\rm \ powers of $L$ down from leading term\ } \as^3 L^6 ; 
\nonumber\\
  \as h^2  \approx  \as^3 L^2, \hspace{2em} &&
4  \mbox{\rm \ powers of $L$ down from leading term}; 
\nonumber\\
  \as h  \approx  \as^2 L , \hspace{2em} &&
3  \mbox{\rm \ powers of $L$ down from leading term\ } \as^2 L^4. 
  \label{eq:Iexpand1}
\end{eqnarray}
Furthermore, the scale choice in the pdfs, $\qt \exp {K}/{2}$, induces a
change of relative order
\begin{equation}
  \as K \approx \as h^2 \approx \as^3 L^2 .
\end{equation}
If we are only interested in terms of relative order $\as^2$ with respect to
the Born contribution, the only term that we need to keep is the one of order $\as
h$, i.e.~the term
\begin{equation}
  - \as \, f''_0 \left( y_0 \right) \, b_0^2 \, \zeta_3 \, h \approx  2 \,\zeta_3\,
  A_1^2 \, \as^2\,  \log \frac{\qt^2}{c_2^2 Q^2} \,.
\end{equation}
This is the term kept by Ellis and Veseli in ref.~\cite{Ellis:1997ii}, and
the replacement
\begin{equation}
 \label{eq:newb2}
  B_2 \rightarrow B_2 + 2 \, \zeta_3 \, A_1^2
\end{equation}
is sufficient to incorporate this term.

\section{Scale variation in the resummed expression}
\label{app:scalevar}
We consider the resummed cross section written in the process-independent form
of eq.~(13) of ref.~\cite{Catani:2000vq}
\begin{equation}
  \frac{\mathd \sigma}{ \mathd \qT^2 \mathd \yB} = \sigma_F \int \mathd^2 b
  \, e^{i \vec{q}_{\sss\rm T} \cdot \vec{b}} \left[ C_{i a} \otimes
    f_{a/A}\right] \left( x_A, \frac{c_1}{b} \right) \left[ C_{j b} \otimes
    f_{b/B} \right] \left( x_B, \frac{c_1}{b} \right) \, \exp \mathcal{S}
    \left( c_2 Q, \frac{c_1}{b} \right),
\end{equation}
where
\begin{equation}
\label{eq:def_H}
  \sigma_F = \sigma_0 \times H    ,
  \hspace{2em} H = 1 + \as(Q^2) H_1+ \ldots 
\end{equation}
We further rewrite this formula as
\begin{eqnarray}
\label{eq:procind} 
  \frac{\mathd \sigma}{\mathd \qT^2 \mathd \yB } & = & \sigma_F \int \mathd^2
  b \, e^{i \vec{q}_{\sss\rm T}\cdot \vec{b}} \left[ C_{ia} \left(
    \frac{c_1}{b}, \mu_F \right) \otimes f_{a/A} \left( \mu_F \right)
    \right] \left( x_A \right) \nonumber\\
& & \phantom{ \sigma_F \int \mathd^2 b \, e^{i \vec{q}_{\sss\rm T}\cdot \vec{b}}}
 \hspace{-4mm}\times \left[ C_{j b} \left(
    \frac{c_1}{b}, \mu_F \right) \otimes f_{b/B} \left( \mu_F \right) \right]
    \left( x_B \right) \, \exp \mathcal{S} \left( c_2 Q, \frac{c_1}{b} \right)
    .
\end{eqnarray}
Since eq.~(\ref{eq:procind}) is process independent, it must be possible to
introduce the scale variations independently in the integrand and in the
prefactor $\sigma_F$. Furthermore, in the prefactors involving the pdfs there
is a built-in scale independence as far as the factorization scale variations
are concerned. Thus, the Sudakov exponent must be written in a
scale-invariant form by itself to recover the full scale dependence.  We thus
write
\begin{equation}
  S \left( c_2 Q,  \frac{c_1}{b} \right) = \exp \left\{ - \int_{c_1^2 / b^2}^{c_2 ^2 Q^2}
  \frac{\mathd \mu^2}{\mu^2} \left[ A \left( \as \left( \mu^2 \right)
  \right) \log \frac{c_2^2 Q^2}{\mu^2} + B \left( \as \left( \mu^2
  \right) \right) \right] \right\} .
\end{equation}
Performing the change of variable
\begin{equation}
  \mu' = \KRA \mu ,
\end{equation}
we get
\begin{equation}
 S \left( c_2 Q,  \frac{c_1}{b} \right) = \exp \left\{ - \int_{\KRA^2 c_1^2 /
  b^2}^{\KRA^2 c_2^2 Q^2} \frac{\mathd \mu'^2}{\mu'^2} \left[ A \left( \as
  \left( \frac{\mu'^2}{\KRA^2} \right) \right) \log \frac{\KRA^2 c_2^2
  Q^2}{\mu'^2} + B \left( \as \left( \frac{\mu'^2}{\KRA^2} \right) \right)
  \right] \right\} .
\end{equation}
We relabel $\mu' \rightarrow \mu$, and, using
\begin{equation}
\as \left( \mu^2 / \KRA^2 \right) = \as \left( \mu^2 \right) -
   b_0 \log \frac{1}{\KRA^2} \as^2 \left( \mu^2 \right) \ldots ,
\end{equation}
we find
\begin{eqnarray}
  A \left( \as \left( \frac{\mu^2}{ \KRA^2} \right) \right) & = & A_1 \as
  \left( \mu^2 \right) + \left( A_2 + 2 b_0 A_1 \log \KRA \right) \as^2
  \left( \mu^2 \right) 
\nonumber\\
  & & +\left[ A_3 + A_1 b_0^2 \left( 2 b_1 \log \KRA + 4 \log^2 \KRA
  \right) + 4 b_0 A_2 \log \KRA \right] \as^3 \left( \mu^2 \right) +
\ord{\as^4}\,, 
\nonumber \\
  B \left( \as \left( \frac{\mu^2 }{ \KRA^2 }\right) \right) & = & B_1 \as
  \left( \mu^2 \right) + \left( B_2 + 2 b_0 B_1 \log \KRA \right) \as^2
  \left( \mu^2 \right) +\ord{\as^3}\,. 
\end{eqnarray}
Defining
\begin{eqnarray}
  A \left( \as \left( \mu^2 \right), \KRA \right) & = & A_1 \as
  \left( \mu^2 \right) + \left( A_2 + 2 b_0 A_1 \log \KRA \right) \as^2
  \left( \mu^2 \right)
\nonumber \\
  & & + \left[ A_3 + A_1 b_0^2 \left( 2 b_1 \log \KRA + 4 \log^2
  \KRA \right) + 4 b_0 A_2 \log \KRA \right] \as^3 \left( \mu^2
  \right)+\ord{\as^4}\,, 
\nonumber\\
  B \left( \as \left( \mu^2 \right), \KRA \right) & = & B_1 \as
  \left( \mu^2 \right) + \left( B_2 + 2 b_0 B_1 \log \KRA \right) \as^2
  \left( \mu^2 \right)
\nonumber \\
  &  & + \left[ A_1 \as \left( \mu^2 \right) + \left( A_2 + 2 b_0 A_1
  \log \KRA \right) \as^2 \left( \mu^2 \right) \right] 2 \log \KRA
\nonumber \\
  & = & \left[ B_1 +2  A_1 \log \KRA  \right]\as \left( \mu^2 \right) 
\nonumber \\
&& + \left[ B_2 + 2\( A_2 + b_0 B_1 \) \log\KRA 
+ 4 b_0 A_1 \log^2 \KRA \right] \as^2  \left( \mu^2 \right)+\ord{\as^3}\,, 
\end{eqnarray}
we get
\begin{equation}
\label{eq:exp_sud_resc}
  S \left( c_2 Q,  \frac{c_1}{b} \right)  =  \exp \left\{ -\!\! \int_{\KRA^2 c_1^2 /
  b^2}^{\KRA^2 c_2^2 Q^2} \!\!  \frac{\mathd \mu^2}{\mu^2} \!\! \left[ A \left( \as \left(
  \left( \mu^2 \right), \KRA \right) \right) \log \frac{c_2^2 Q^2}{\mu^2} +
  B \left( \as \left( \mu^2 \right), \KRA \right) \right] \!\right\}.
\end{equation}
We can now break the integral in the exponent into two integrals
\begin{equation}
\label{eq:break_int}
- \int_{\KRA^2 c_1^2 /  b^2}^{\KRA^2 c_2^2 Q^2} \ldots = 
-  \int_{\KRA^2 c_1^2/b^2}^{c_2^2 Q^2}\ldots  - \int_{c_2^2 Q^2}^{\KRA^2 c_2^2 Q^2} \ldots\,,
\end{equation}
and we call the last integral $I$
\begin{equation}
\label{eq:def_I}
I \equiv  - \int_{c_2^2 Q^2}^{\KRA^2 c_2^2 Q^2} \frac{\mathd \mu^2}{\mu^2} \left[ A \left(
  \as \left( \left( \mu^2 \right), \KRA \right) \right) \log
  \frac{c_2^2 Q^2}{\mu^2} + B \left( \as \left( \mu^2 \right), \KRA
  \right) \right].
\end{equation}
Since this integral does not have large logarithms, we can evaluate it at order
$\as \left( c_2^2 Q^2 \right)$, ignoring the running of $\as$,
\begin{eqnarray}
\label{eq:int_to_add}
 I &=& A_1 \as \left( c_2^2 Q^2 \right) \int_{\KRA^2 c_2^2 Q^2}^{c_2^2 Q^2}
  \frac{\mathd \mu^2}{\mu^2} \log \frac{c_2^2 Q^2}{\mu^2} + \left( B_1 + 2 A_1 
  \log \KRA \right) \as \left( c_2^2 Q^2 \right) \int_{\KRA^2 c_2^2 Q^2}^{c_2^2
  Q^2} \frac{\mathd \mu^2}{\mu^2} 
\nonumber\\
   &=& \as \left( c_2^2 Q^2 \right) \left[ 2 A_1  \log^2 \KRA - 2\left(
  B_1 + 2 A_1 \log \KRA \right) \log \KRA \right] 
\nonumber\\
  &=& - 2\, \as \left( c_2^2 Q^2 \right) \left[  A_1 \log^2 \KRA + B_1 \log\KRA \right] . 
\end{eqnarray}
Using the renormalization group equation for $\as$
\begin{equation}
\mu^2 \frac{\mathd \as}{\mathd \mu^2} = - b_0 \as^2 \qquad\Longrightarrow\qquad
\as^2 \,\frac{\mathd \mu^2}{\mu^2} = -\frac{1}{b_0}\mathd \as\,,
\end{equation}
we can write the identity
\begin{equation}
 \label{eq:b2evol}
  - \int_{\KRA^2 c_1^2 / b^2}^{c_2^2 Q^2} \frac{\mathd \mu^2}{\mu^2} \as^2
  \left( \mu^2 \right) =  \int_{\KRA^2 c_1^2 / b^2}^{c_2^2 Q^2}
  \frac{1}{b_0} \mathd \as \left( \mu^2 \right) = \frac{1}{b_0} \left[ \as
    \left( c_2^2 Q^2 \right) - \as \left(  \frac{\KRA^2\, c_1^2}{ b^2} \right) \right]\,.
\end{equation}
Solving this equation for $ \as\left( c_2^2 Q^2 \right)$ and inserting it
into the expression of $I$ of eq.~(\ref{eq:int_to_add}) we get
\begin{eqnarray}
\label{eq:final_I}
 I &=&   -\int_{\KRA^2 c_1^2 / b^2}^{c_2^2 Q^2} \frac{\mathd \mu^2}{\mu^2} \,\as^2 \left(
  \mu^2 \right)  \left[ -2 b_0 A_1  \log^2 \KRA - 2 b_0 B_1 \log \KRA
    \right]
\nonumber\\
&&  - \, \as \left(  \frac{\KRA^2\,c_1^2}{ b^2} \right) \left[2 A_1 \log^2 \KRA +
 2 B_1  \log \KRA \right].
\end{eqnarray}
Using eqs.~(\ref{eq:break_int}), (\ref{eq:def_I}) and~(\ref{eq:final_I}),
we can write eq.~(\ref{eq:exp_sud_resc}) as
\begin{eqnarray}
S \left( c_2 Q, \frac{c_1}{b} \right) & = & \exp \left\{ - \int_{\KRA^2 c_1^2 /
    b^2}^{c_2^2 Q^2} \frac{\mathd \mu^2}{\mu^2} \left[ A \left( \as \left(
    \left( \mu^2 \right), \KRA \right) \right) \log \frac{c_2^2 Q^2}{\mu^2}
    + B' \left( \as \left( \mu^2 \right), \KRA \right) \right]    \right\}
\nonumber \\
    & &\times \left\{ 1 - \as \left( \frac{\KRA^2 c_1^2}{b^2} \right) \left[
   2  A_1 \log^2 \KRA + 2 B_1  \log \KRA \right] \right\},
\label{eq:corrected}
\end{eqnarray}
where
\begin{equation}
\label{eq:Bprime}
  B'\!\left( \as \left( \mu^2 \right), \KRA \right) = \left[ B_1 +2 A_1 \log
    \KRA \right]\as \!\left( \mu^2 \right) + \left[ B_2 + 2 A_2\log\KRA + 2\,
    b_0 A_1 \log^2 \KRA \right] \!\as^2\! \left( \mu^2 \right).
\end{equation}
The last term in curly braces of eq.~(\ref{eq:corrected}) affects the
$C_{ij}$ terms by a contact term. In our case it is irrelevant, since we
always deal with the $\mathcal{O} \left( \as^2 \right)$ expanded cross
section multiplied by the Sudakov exponential.  Notice also that the argument
of $\as$ in the contact term can be changed by a factor of order 1 at the
required accuracy. Thus, the $\KRA$ dependence drops from there, and $\as$
can be evaluated at a scale of the order of the low scale, i.e.~$c_1^2/b^2$.

\subsection{Process dependent form}
We can also translate eq.~(\ref{eq:Bprime}) to the case of a
process-dependent form, which is similar to formula~(\ref{eq:procind}) except
that the prefactor $\sigma_F$ is replaced by $\sigma_0$. In order to perform
this translation, the $H$ factor must be somehow incorporated in the $C_{ij}$
coefficients and in the Sudakov exponent. In order to do this, we first
notice that if the Born term is of order $\as^n$ in the strong coupling
constant, the scale dependence of $H$ is derived using the identity
\begin{equation}
  \as^n \left( \KRA^2 Q^2 \right) H_c^F \left( \as \left( \KRA^2
  Q^2 \right), \KRA \right) = \as^n \left( Q^2 \right) H_c^F \left(
  \as \left( Q^2 \right), 1 \right),
\end{equation}
which, using eq.~(\ref{eq:def_H}), leads to
\begin{equation}
 H_c^F \left( \as \left( \KRA^2 Q^2 \right), \KRA \right) = 1 +
   \as \left( \KRA^2 Q^2 \right) \left[ H_1 + n b_0 \log \KRA^2 \right]+\dots
 \end{equation}
This can be replaced by
\begin{eqnarray}
  H_c^F \left( \as \left( \KRA^2 Q^2 \right), \KRA \right) &=& \left[ 1 +
  \as \left( \frac{\KRA^2\, c_1^2}{b^2} \right) \left( H_1 + n b_0 \log \KRA^2
  \right) \right] 
\nonumber\\
&&{}\times
\exp \left\{ (H_1 + n b_0 \log \KRA^2) \left[\as \left(\KRA^2 Q^2 \right) - \as
  \left( \frac{\KRA^2\, c_1^2}{b^2} \right) \right] \right\}\!, \phantom{aaaa}
\end{eqnarray}
and using eq.~(\ref{eq:b2evol})
\begin{eqnarray}
\label{eq:Heffe} 
  H_c^F \left( \as \left( \KRA^2 Q^2 \right), \KRA \right) & = & \left[
  1 + \as \left(  \frac{\KRA^2\, c_1^2}{b^2}\right) \left( H_1 +2 n b_0 \log \KRA
  \right) \right] \nonumber\\
  & & {}\times  \exp \left\{ - \int_{\KRA^2 c_1^2 / b^2}^{\KRA^2 Q^2} \frac{\mathd
  \mu^2}{\mu^2} \as^2 \left( \mu^2 \right) b_0 \left[ H_1 + 2 n b_0 \log
  \KRA \right] \right\}.\phantom{aaaa}
\end{eqnarray}
On the other hand, the scale dependence in the resummation factor must be
equal to what we have found in eq.~(\ref{eq:corrected}). Inserting
eq.~(\ref{eq:Heffe}) in the process independent formula, we should get the
process dependent result. This induces the modification\footnote{Note the
  difference of a factor 2 between the last term in eq.~(\ref{eq:B2_F}) and
  the corresponding factor in eq.~(\ref{eq:B2tilde}), due to the square
  factor in the definition of the Sudakov form factor we are using in
  eq.~(\ref{eq:calSdef}).}
\begin{equation}
\label{eq:B2_F}
  B_2^F = B_2 + b_0 H_1 +2 n b_0^2 \log \KRA,
\end{equation}
and the modification of the $C_{ij}$ coefficients by a contact term
\begin{equation}
  C_{ij} \left( z \right) \rightarrow C_{ij} \left( z \right) + \delta
  \left( 1 - z \right)  \frac{1}{2} \left( H_1 +2 n b_0 \log \KRA \right)
  \as \left( \frac{\KRA^2\, c_1^2}{b^2} \right) ,
\end{equation}
where the $1/2$ factor comes from the fact that, in the resummed expressions,
there is the product of two $C_{ij}$ terms. 

\section{A mathematical complement}
\label{sec:integrals}
In this section, we explicitly estimate the size of the following integral,
that we use throughout to estimate the contributions to the inclusive cross
section
\begin{equation}
\label{eq:Imn}
I(m,n) \equiv \int^{Q^2}_{\Lambda^2} \frac{\mathd q^2}{q^2} \(\log
\frac{Q^2}{q^2}\)^m \as^n \left(q^2\right) \exp\lg -\int_{q^2}^{Q^2} 
\frac{\mathd \mu^2}{\mu^2} A \, \as\(\mu^2\)  \log\frac{Q^2}{\mu^2} \rg\,,
\end{equation}
with
\begin{equation}
\as(\mu^2) = \frac{1}{b_0 \log\frac{\mu^2}{\Lambda^2}}\,,
\end{equation}
where $\Lambda$ is the usual $\LambdaQCD$. We first evaluate the argument of
the exponent
\begin{eqnarray}
\int_{q^2}^{Q^2} 
\frac{\mathd \mu^2}{\mu^2} A \, \as\(\mu^2\)  \log\frac{Q^2}{\mu^2} &=&
\int^{Q^2}_{q^2} \mathd \( \log\frac{\mu^2}{\Lambda^2}\)
 A \frac{1}{b_0 \log\frac{\mu^2}{\Lambda^2}} 
  \left[ \log\frac{Q^2}{\Lambda^2}-\log\frac{\mu^2}{\Lambda^2}\right]
\nonumber\\
&=& \frac{A}{b_0} 
\int^{L}_{l} \mathd x \frac{1}{x}  \( L - x \)
=   \frac{A}{b_0}  \( L \log\frac{L}{l} - L + l\)\,,
\end{eqnarray}
where we have defined
\begin{equation}
 l = \log\frac{q^2}{\Lambda^2}\,, \qquad \qquad 
 L = \log\frac{Q^2}{\Lambda^2}\,, \qquad \qquad 
 x = \log\frac{\mu^2}{\Lambda^2}\,.
\end{equation}
Equation~(\ref{eq:Imn}) then becomes
\begin{eqnarray}
I(m,n) &=& \int^{L}_{0} \mathd l \, \(L-l\)^m
 \frac{1}{b_0^n \, l^n} \exp\lg 
-a \lq  L \( \log L -  \log l\) -  L + l\rq \rg
\nonumber\\
&=& \frac{1}{b_0^n}
\int^{L}_{0} \mathd l \, \exp\lg
m\log(L-l) -n\log l 
-a \lq  L \( \log L -  \log l\) -  L + l\rq \rg
\nonumber\\
&=& \frac{1}{b_0^n}
\int^{L}_{0} \mathd l \, \exp \lq f(l) \rq \,,
\end{eqnarray}
where 
\begin{equation}
f(l) = m\log(L-l) -n\log l -a \lq  L \( \log L -  \log l\) -  L + l \rq\,,
\end{equation}
and we have defined $a=A/b_0$. This integral has to be computed for large
$L$.  We look for an approximation of the integral using  the  saddle-point
technique, i.e.~we expand the argument of the exponent around its maximum
\begin{equation}
f(l) = f(l_M) + \frac{1}{2}\, f''(l_M) \,(l-l_M)^2 + \ord{\(l-l_M\)^3}
\end{equation}
with
\begin{equation}
l_M= \frac{1}{2a} \lq m - n + 2 \, a\, L - \sqrt{(m-n)^2 + 4 \,a \,m \,L} \rq
=L - \sqrt{\frac{m L }{a}} + \ord{1} \,.
\end{equation}
For large $L$ we have
\begin{eqnarray}
f(l_M) &=& \( \frac{m}{2} -n \) \log L + \ord{1}\,,
\\
f''(l_M) &=& -  \frac{2a}{L} + \ord{L^{-3/2}}\,,
\end{eqnarray}
and we get
\begin{eqnarray}
I(m,n) &\approx& \frac{1}{b_0^n} \, L^{\frac{m}{2} -n} \int_0^L \mathd l
\exp \lq -\frac{a}{L}(l-l_M)^2 \rq =
\frac{1}{b_0^n} \, L^{\frac{m}{2} -n} \int_{-l_M}^{L-l_M} \mathd l'
\exp \lq -\frac{a}{L} l'^2 \rq
\nonumber\\
&=&\frac{1}{b_0^n} \frac{1}{\sqrt{a}}
\, L^{\frac{m+1}{2} -n} \int_{-\sqrt{\frac{a}{L}} l_M}^{\sqrt{\frac{a}{L}} (L-l_M)} \mathd x
\exp \( -x^2\) 
\approx \frac{1}{b_0^n} \frac{1}{\sqrt{a}}
\, L^{\frac{m+1}{2} -n} \int_{-\sqrt{aL}}^{\sqrt{m}} \mathd x
\exp \( -x^2\) 
\nonumber\\
&\le& \frac{1}{b_0^n} \frac{1}{\sqrt{a}}
\, L^{\frac{m+1}{2} -n} \int_{-\infty}^{\infty} \mathd x
\exp \( -x^2\) = \frac{1}{b_0^n} \,\sqrt{\frac{\pi}{a}}
\, L^{\frac{m+1}{2} -n} \approx \lq \as(Q^2) \rq^{n - \frac{m+1}{2}}\,.
\end{eqnarray}
In other words, for each power of the logarithm in eq.~(\ref{eq:Imn}),
we lose half power of $\as$.

\providecommand{\href}[2]{#2}\begingroup\raggedright\endgroup

\end{document}